\documentclass[preprints,article,accept,moreauthors,pdftex,10pt,a4paper]{Definitions/mdpi} 
\firstpage{1} 
\pubvolume{xx}
\issuenum{1}
\articlenumber{5}
\pubyear{2019}
\copyrightyear{2019}
\history{Received: date; Accepted: date; Published: date}
\usepackage[colorinlistoftodos]{todonotes}
\newcommand{\RN}[1]{\textup{\uppercase\expandafter{\romannumeral#1}}}

\pdfoutput=1

\usepackage{color}
\usepackage{epsfig}
\usepackage{amsmath}
\usepackage{amssymb}
\usepackage{bm}
\usepackage{graphicx}
\graphicspath{ {images/} }
\usepackage{float}
\usepackage{mathtools}
\usepackage{comment}

\Title{Nonlocal Scalar Quantum Field Theory: Functional Integration, Basis Functions Representation and Strong Coupling Expansion}

\Author{Matthew Bernard $^{1,}$*, Vladislav A. Guskov $^{2}$, Mikhail G. Ivanov $^{2,}$*, Alexey E. Kalugin $^{2}$ and Stanislav L. Ogarkov $^{2,}$*}

\DeclarePairedDelimiter{\norm}{\lVert}{\rVert\,}

\AuthorNames{Matthew Bernard, Vladislav A. Guskov, Mikhail G. Ivanov, Alexey E. Kalugin and Stanislav L. Ogarkov}

\address{%
$^{1}$ \quad Advanced Computational Biology Center (ACBC), University Avenue 2120, 94704 Berkeley, CA, USA;\\
$^{2}$ \quad Moscow Institute of Physics and Technology (MIPT), Institutskiy Pereulok 9, 141701 Dolgoprudny, \quad Moscow Region, Russia; guskov.va@phystech.edu (V.A.G.); kalugin.ae@phystech.edu (A.E.K.);}

\corres{Correspondence: mattb@berkeley.edu (M.B.); ivanov.mg@mipt.ru (M.G.I.); ogarkovstas@mail.ru (S.L.O.)}

\abstract{Nonlocal quantum field theory (QFT) of one-component scalar field $\varphi$ in $D$-dimensional Euclidean spacetime is considered. The generating functional (GF) of complete Green functions $\mathcal{Z}$ as a functional of external source $j$, coupling constant $g$, and spatial measure $d\mu$ is studied. An expression for GF $\mathcal{Z}$ in terms of the abstract integral over the primary field $\varphi$ is given. An expression for GF $\mathcal{Z}$ in terms of integrals over the primary field and separable Hilbert space (HS) is obtained by means of a separable expansion of the free theory inverse propagator $\hat{L}$ over the separable HS basis. The classification of functional integration measures $\mathcal{D}\left[\varphi\right]$ is formulated, according to which trivial and two nontrivial versions of GF $\mathcal{Z}$ are obtained. Nontrivial versions of GF $\mathcal{Z}$ are expressed in terms of $1$-norm and $0$-norm, respectively. In the $1$-norm case in terms of the original symbol for the product integral, the definition for the functional integration measure $\mathcal{D}\left[\varphi\right]$ over the primary field is suggested. In the $0$-norm case, the definition and \emph{the meaning} of $0$-norm are given in terms of the replica-functional Taylor series. The definition of the $0$-norm generator $\varPsi$ is suggested. Simple cases of sharp and smooth generators are considered. An alternative derivation of GF $\mathcal{Z}$ in terms of $0$-norm is also given. All these definitions allow to calculate corresponding functional integrals over $\varphi$ in quadratures. Expressions for GF $\mathcal{Z}$ in terms of integrals over the separable HS, aka the basis functions representation, with new integrands are obtained. For polynomial theories $\varphi^{2n},\, n=2,3,4,\ldots,$ and for the nonpolynomial theory $\sinh^{4}\varphi$, integrals over the separable HS in terms of a power series over the inverse coupling constant $1/\sqrt{g}$ for both norms ($1$-norm and $0$-norm) are calculated. Thus, the strong coupling expansion in all theories considered is given. ``Phase transitions'' and critical values of model parameters are found numerically. A generalization of the theory to the case of the uncountable integral over HS is formulated: GF $\mathcal{Z}$ for an arbitrary QFT and the strong coupling expansion for the theory $\varphi^{4}$ are derived. Finally a comparison of two GFs $\mathcal{Z}$, one on the continuous lattice of functions and one obtained using the Parseval--Plancherel identity, is given.}


\newcommand{\la}{\langle}
\newcommand{\ra}{\rangle}
\newcommand{\lp}{\left(}
\newcommand{\rp}{\right)}
\newcommand{\lc}{\left\{}
\newcommand{\rc}{\right\}}
\newcommand{\lb}{\left[}
\newcommand{\rb}{\right]}
\usepackage{comment}

\newcommand*{\prodint}{\raisebox{-14 pt}{\includegraphics[height=2.79em]{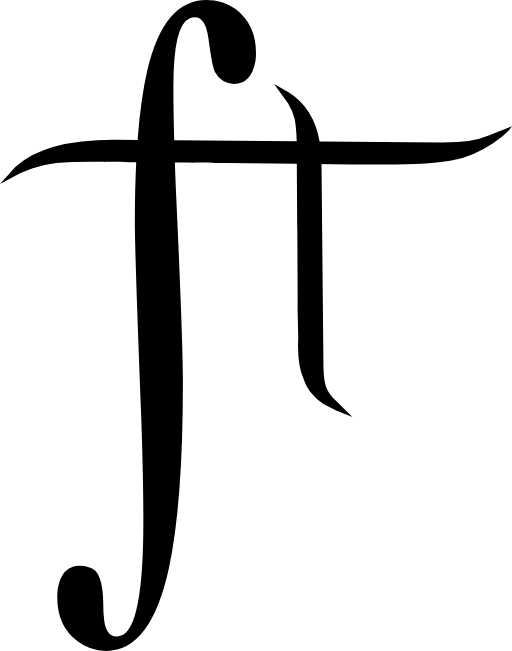}}}
\newcommand*{\bind}{\!\!\!\!\!\!\!\!}

\begin{document}

\section{Introduction}

Calculation of functional (path) integrals is an important problem in quantum field theory (QFT) \cite{bogoliubov1980introduction,bogolyubov1983quantum,bogolyubov1990GeneralQFT} as well as in other fields of theoretical and mathematical physics and infinite-dimensional analysis \cite{vasil2004field,zinn1989field,Mazzucchi2008,Mazzucchi2009,DeWitt-Morette,Montvay,Kleinert,Johnson,Steiner,Mosel,Simon,Popov 1976,Shavgulidze2015}. An expansion over a small coupling constant $g$, aka a perturbation theory (PT) over $g$, does not solve the problem: Strong coupling limit must be known (some modifications of PT are described in \cite{BelokurovI,BelokurovII,korsun1992VPT}). Moreover, a behavior of generating functionals (GFs) of different Green functions families is unknown for large coupling constants $g$ beyond the PT for most of interaction Lagrangians.

In this paper, we consider two types of a nonlocal QFT of a one-component scalar field in $D$-dimensional Euclidean spacetime: polynomial and nonpolynomial \cite{efimov1970nonlocal,alebastrov1973proof,alebastrov1974proof,efimov1977cmp,efimov1979cmp,efimov1977nonlocal,efimov1985problems,efimov2001amplitudes,efimov2004blokhintsev,basuev1973conv,basuev1975convYuk,Ogarkov2019,Kosyakov2019}. A scalar field theory underlies multiple physical models: Higgs Scalar Particles, Theory of Critical Phenomena (note an interesting book \cite{nedelko1995oscillator}), Quantum Theory of Magnetism, Plasma Physics (note an interesting review \cite{brydges1999review}), Theory of Developed Turbulence, Nuclear Physics (note an interesting book \cite{ivanov1993quark} and some interesting papers \cite{Ganbold2019,Gutsche2019}), etc. Being the simplest theory, it serves as a ``building block'' of more complicated physical models.

We formulate GF $\mathcal{Z}$ of complete Green functions as a functional of external source $j$, coupling constant $g$, and a spatial measure $d\mu$ in terms of the abstract (formal) functional integral over primary field $\varphi$, as a formal variable \cite{kopbarsch,wipf2012statistical}. Using a separable expansion of a free theory inverse propagator $\hat{L}$ over a separable Hilbert space basis (similar to \cite{Ogarkov2019}, where the expansion of the propagator $\hat{G}$ itself was used), we derive an expression for GF $\mathcal{Z}$ in terms of a ``double'' integral: over the primary field $\varphi$ and the separable HS (infinite and countable lattice of auxiliary variables $t_{s}$).

Results of further calculations drastically depend on the definition of the functional integration measure $\mathcal{D}\left[\varphi\right]$. First, we remind a definition of measure $\mathcal{D}\left[\varphi\right]$ such that the expression for GF $\mathcal{Z}$ turns out to be trivial. Then we formulate two definitions of measure $\mathcal{D}\left[\varphi\right]$ for which GF $\mathcal{Z}$ is similar to that for a lattice QFT and spin models or their deformations. These nontrivial versions of GF $\mathcal{Z}$ are expressed in terms of $1$-norm and $0$-norm, respectively. 

In the first case we suggest an original symbol for the product integral formulation of functional integration measure $\mathcal{D}\left[\varphi\right]$ definition. Such a symbol is defined not only over the set of spatial coordinates $X$, but also over the set of spatial measures $d\mu(x)$. It is convenient for lattice QFT type calculations. In the second case, we suggest the definition of $0$-norm in terms of the replica-functional Taylor series. This definition demonstrates the meaning of $0$-norm: The norm $\,\norm{f}_{0}$ is defined not axiomatically, but as some function of distance in the space of functions between the considered one $f$ and identically zero function (the distance from the function to the origin of coordinates in the corresponding space of functions). The definition is sufficient for the purposes of this paper.

A special case of $0$-norm definition is the one in terms of a function $\varPsi$ that generates this norm. The function $\varPsi$ is therefore called $0$-norm generator. We consider one sharp generator $\varPsi$ and two smooth ones. In the first case, $0$-norm and $1$-norm are equivalent. The second case reproduces smooth deformation: An allegory of the functional (nonperturbative) renormalization group (FRG) method \cite{kopbarsch,wipf2012statistical}, in which FRG flow regulators can be sharp or smooth functions. Trigonometric $\arctan(z)$ and hyperbolic $\tanh(z)$ functions are chosen as two smooth generators. We note that chosen models are used for illustrative purposes only. In a real physical model, such generators are part of the problem statement and are determined by physical requirements. Being an additional degree of freedom, such a description is useful for problems of the Nuclear Physics and effective field theories in the Quantum Chromodynamics (QCD) and the Quantum Electrodynamics (QED).

In principle mathematical technique developed allows to calculate the functional integral over $\mathcal{D}\left[\varphi\right]$ in quadratures for any nonlocal theories considered in this paper. The remaining integral over HS is the generating functional $\mathcal{Z}$ in basis functions representation. We note that in this case an integrand is a more complicated expression than initial exponential function of action. GF $\mathcal{Z}$ in a basis functions representation is open to numerical methods for QFT on lattice and spin models with rather efficient computational use of resources, as an alternative to majorize GF $\mathcal{Z}$ or the scattering matrix ($\mathcal{S}$-matrix) within a framework of variational principle: solving corresponding variational problem to obtain a physically satisfactory estimate for $\mathcal{Z}$ or $\mathcal{S}$ \cite{efimov1985problems,Ogarkov2019}.

To demonstrate the method, outlined in the present paper, we suggest to consider generally accepted models: polynomial theories $\varphi^{2n},\, n=2,3,4,\ldots,$ and the nonpolynomial theory $\sinh^{4}\varphi$ in $D$-dimensional Euclidean spacetime. The calculation of GF $\mathcal{Z}$ can be done analytically in terms of a power series over the inverse coupling constant $1/\sqrt{g}$, aka the strong coupling expansion, for all types of norms discussed in the paper. We note that such an expansion resembles the hopping parameter expansion in a lattice QFT as well as a high-temperature expansion in statistical physics \cite{wipf2012statistical}. However, results turn out to be nontrivial for any dimension $D$ of spacetime.

We verify obtained results by using a nonlocal inverse propagator $\hat{L}$. Although, if we consider quasilocal $\hat{L}$, an additional renormalization scheme is necessary, but it does not affect convergence properties of series over $1/\sqrt{g}$ as a whole. Thus, we propose a broad mathematical technique that allows to go beyond a PT and is able to shed light on a nonperturbative physics, described by various scalar field theories. The development of the presented approach is useful for the low-energy QCD and high-energy QED physics description. In the case of a sharp generator $\varPsi$ (nonpolynomial theory $\sinh^{4}\varphi$) and in the case of a smooth generator $\varPsi$ (polynomial and nonpolynomial theories), the $1/\sqrt{g}$ correction is alternating. Thus, the considered models can describe phase transitions, critical values of which (in the simplest case) are determined from zero value of $1/\sqrt{g}$ correction. We present numerical values of critical parameters for two QFTs $\varphi^{4}$ and $\sinh^{4}\varphi$ and two $0$-norms with generators $\arctan(z)$ and $\tanh(z)$.

For a better understanding of the mathematical technique, developed in our paper, we give a generalization of the theory to the case of the uncountable integral over the HS. First, the general theory is considered and then calculations are performed for the polynomial theory $\varphi^{4}$. The results of this generalization coincide with the analogous ones in the case of the countable integral over HS. Finally, we derive an expression for GF $\mathcal{Z}$, starting from the Parseval--Plancherel identity (page $176$ in \cite{Shavgulidze2015}). The result obtained repeats the expression for GF $\mathcal{Z}$ in terms of the uncountable integral over HS. This coincidence, being an independent verification method, shows the correctness of the theory developed in the paper.

Since nonlocal theories study is motivated, in particular, by attempts to construct the quantum theory of gravity (QG) \cite{moffat2011gravity,moffat2016,Modesto2014,Modesto2018} and QFT in a curved spacetime, we would like to point out that the basis functions representation can draw parallels between QG in continuous spacetime and QG on the discrete lattice in HS. QG theory of this kind is popular nowadays, in particular, as Loop QG. Also note a paper on a study of nonlocality in string theory \cite{Modesto2015}.

This paper has the following structure: In section $2$ we propose a general theory, sections $3$ and $4$ are devoted to the polynomial and nonpolynomial QFTs, respectively. Section $5$ is devoted to QFT on the continuous lattice of functions (uncountable HS integral) and the Parseval--Plancherel identity. In the conclusions, section $6$, we give a final discussion of all results obtained in the paper, and highlight further possible areas of research.

\section{General Theory}
\vspace{-6pt}
\subsection{Derivation of GF $\mathcal{Z}$ in Basis Functions Representation: Trivial Integration}
\vspace{-6pt}
\subsubsection{Inverse Propagator Splitting}

Traditionally we begin with a GF of complete Green functions $\mathcal{Z}$ in terms of an abstract integral over primary field $\varphi$ \cite{efimov1977nonlocal,efimov1985problems,Ogarkov2019,kopbarsch,wipf2012statistical}:
\begin{equation}\label{generation_function_interaction}
    \mathcal{Z}[j,g]=\int \mathcal{D}[\varphi]\, e^{-S_{0}[\varphi]-S_{1}[g,\varphi] + (j|\varphi)},
\end{equation}
where $S_{0}$ is a free theory action, $S_{1}$ is an interaction action, $g$ and $j$ are functions, a coupling constant and a source, that we feed to our functional. As for interaction action ($\varphi_{0}$ is a field scale)
\begin{equation}
    S_{1}\lb g,\varphi\rb = \int d^{D}x g(x) U \lb \frac{\varphi(x)}{\varphi_{0}}\rb,
\end{equation}
theories with interaction of the form $U\left(\varphi\right) = \varphi^{2n},\, \sinh^{2n}\varphi,\,n=2,3,4,\ldots,$ etc. can be considered. In the presented general theory it is important that $U\left(-\varphi\right)=U\left(\varphi\right)$ and $U\left(\varphi\rightarrow\infty\right)\rightarrow\infty$. Let's keep things general for now and move forward in the derivation of a theory without specific interaction as far as possible. We will specify a theory latter on to demonstrate a developed technique. Taking a closer look on free theory action $S_{0}$ one can rewrite it using basis functions representation (more on basis function representation in \cite{efimov1977cmp,efimov1979cmp,efimov1977nonlocal,efimov1985problems,Ogarkov2019}):
\begin{equation}\label{Inverse_Propagator_Splitting}
\begin{split}
    S_{0}\lb\varphi\rb &= \frac{1}{2} \lp\varphi\left|\hat{L} \right|\varphi\rp = \frac{1}{2} \int d^{D}x \int d^{D}y\, \varphi(x) L(x,y) \varphi(y) \\
    &= \frac{1}{2} \int d^{D}x \int d^{D}y\, \varphi(x) \varphi(y) \sum\limits_{s} D_{s}(x)D_{s}(y) \\
    &= \frac{1}{2} \sum\limits_{s} \lp\int d^{D}x  D_{s}(x)\varphi(x) \rp \lp\int d^{D}y  D_{s}(y)\varphi(y) \rp \\ 
    &=\frac{1}{2}\sum\limits_{s}\lp D_{s}|\varphi \rp^{2}=\frac{1}{2}\lp \vec D|\varphi \rp^{2}.
\end{split}
\end{equation}

The operator $\hat{L}$ is the inverse Gaussian propagator, the function $L(x,y)$ is the kernel of $\hat{L}$. We consider this operator to be nonlocal and expandable in a sense given in the previous expression. For the sake of more convenient notation HS vector notation instead of HS index $s$ is used in the last line of expression (\ref{Inverse_Propagator_Splitting}). An exponent with a free theory action can be rewritten using Gaussian integral trick (Hubbard--Stratonovich transformation) with auxiliary variables $t_{s}$:
\begin{equation}\label{Hubbard_Stratonovich_transformation}
    e^{-S_{0}\lb\varphi\rb} =  \prod\limits_{s}e^{-\frac{1}{2}\lp D_{s}|\varphi\rp^{2}} =\prod\limits_{s}\int\limits_{-\infty}^{+\infty}\frac{dt_{s}}{\sqrt{2\pi}}\, e^{-\frac{1}{2}t_{s}^{2} + it_{s}\lp D_{s}|\varphi\rp}=\int d\sigma_{t}\, e^{\,i\vec{t}\cdot(\vec{D}|\varphi)},
\end{equation}
where the Gaussian (Efimov) measure $d\sigma_{t}$ over HS is introduced:
\begin{equation}
    \int d\sigma_{t} = \prod\limits_{s}\int\limits_{-\infty}^{+\infty}\frac{dt_{s}}{\sqrt{2\pi}}\, e^{-\frac{1}{2}t_{s}^{2}}.
\end{equation}

Everything mentioned gives us GF $\mathcal{Z}$ as (it is important to note that we have rearranged the integration over $d\sigma_{t}$ and $\mathcal{D}[\varphi]$):
\begin{equation}\label{Interaction_exponent}
     \mathcal{Z}[j,g]=\int d\sigma_{t}\int \mathcal{D}[\varphi] \ e^{-S_{1}\lb g,\varphi\rb + (J_{t}|\varphi)},
\end{equation}
where complete source $J_{t} =j + i\vec{t}\cdot\vec{D},$ and its scalar product with a primary field $\varphi$ is:
\begin{equation}
     \lp J_{t}| \varphi \rp= \int d^{D}x J_{t}(x) \varphi(x) = \int d^{D}x j(x) \varphi(x) + i \vec{t}\cdot\int d^{D}x \vec{D}(x)\varphi(x).
\end{equation}

Thus, we obtain an expression for GF $\mathcal{Z}$ as a ``double'' integral: over $\varphi$ and Hilbert space.

\subsubsection{Primary Field Integration}

Remaining exponent in the expression (\ref{Interaction_exponent}) we represent using standard continuous product notation ($X$ is the set of all values $x$):
\begin{equation}
  e^{-S_{1}\lb g,\varphi\rb + (J_{t}|\varphi)} = e^{\,\int d^{D}x \left\{-g(x)U\lb\frac{\varphi(x)}{\varphi_{0}}\rb + J_{t}(x)\varphi(x)\right\}} = \prod\limits_{X}e^{\,d^{D}x \left\{-g(x)U\lb\frac{\varphi(x)}{\varphi_{0}}\rb + J_{t}(x)\varphi(x)\right\}}.
\end{equation}

As already mentioned GF $\mathcal{Z}$ is a functional of a source $j$ and a coupling constant $g$. Further in the paper we will consider more general construction replacing a spatial measure $d^{D}x$ by $d\mu(x)$. After that a GF $\mathcal{Z}$ will become a functional of a measure $d\mu$ as well. A motivation for it is to exclude possibilities to face with divergences of a kind $\int d^{D}x = \infty$ (if there is one). Our new measure $d\mu$ satisfies a normalization condition:
\begin{equation}
    \int d\mu(x) = \mathcal{V}.
\end{equation}

For the normalization of measure we use a certain quantity $\mathcal{V}$, so that it is possible to investigate the dependence on the latter, if necessary. Otherwise one can set $\mathcal{V}=1$.

Now it is time to specify a functional integration measure $\int \mathcal{D}[\varphi]$. Its naive definition:
\begin{equation}\label{Pretty_bad_story}
    \int \mathcal{D}[\varphi]\equiv \prod\limits_{X}\int\limits_{-\infty}^{+\infty}
    \frac{d\varphi(x)}{\varphi_{0}},
\end{equation}
leads to a pretty bad story. Let's show why this is ill-defined and a new definition is needed. Substitute measure (\ref{Pretty_bad_story}) in the expression (\ref{Interaction_exponent}) for GF $\mathcal{Z}$ and with a couple of trivial transformations come to the following form of GF $\mathcal{Z}$:
\begin{equation}\label{GFZ-with_continuous_sum}
\begin{split}
    \mathcal{Z}[j,g;d\mu]&=\int d\sigma_{t}\int\mathcal{D}\lb\varphi\rb e^{\,\int d\mu\lp x\rp\left\{-g(x)U\lb\frac{\varphi(x)}{\varphi_{0}}\rb + J_{t}(x)\varphi(x)\right\}}\\
    &=\int d\sigma_{t}\prod\limits_{X}\exp{\ln\int
    \limits_{-\infty}^{+\infty}
    \frac{d\varphi(x)}{\varphi_{0}}\,
    e^{\,d\mu\lp x\rp f_{j}
    \left[\varphi(x);x\right]}}\\
    &=\int d\sigma_{t}\exp{\int\limits_{X}
    \ln\int\limits_{-\infty}^{+\infty}
    \frac{d\varphi(x)}{\varphi_{0}}\,
    e^{\,d\mu\lp x\rp f_{j}\left[\varphi(x);x\right]}},
\end{split}
\end{equation}
where $f_{j}\lb\varphi;x\rb=-g(x)U\left(\frac{\varphi}
{\varphi_{0}}\right) + J_{t}(x)\varphi$ and the symbol $\int\limits_{X}$ means continuous sum over all $x$. 

Consider the last line of the expression (\ref{GFZ-with_continuous_sum}) in more detail. The expression standing under the sign of a continuous sum can be expanded into series over spatial measure $d\mu$:
\begin{equation}\label{Singular}
    \int\limits_{X}\ln\int\limits_{-\infty}^{+\infty}
    \frac{d\varphi(x)}{\varphi_{0}}\,
    e^{\,d\mu\lp x\rp f_{j}\left[\varphi(x);x\right]}=
    \int\limits_{X}\ln\int\limits_{-\infty}^{+\infty}
    \frac{d\varphi(x)}{\varphi_{0}}\lc 1+d\mu\lp x\rp f_{j}\left[\varphi(x);x\right]+\mathcal{O}\lb\lp d\mu\rp^{2}\rb\rc.
\end{equation}

Expression (\ref{Singular}) needs to be regularized. The regularized version is:
\begin{equation}\label{Trivial_integration_for_Reviewer}
\begin{split}
    &\int\limits_{X}\ln\int\limits_{-\varLambda}^{+\varLambda}
    \frac{d\varphi(x)}{2\varLambda}\,
    e^{\,d\mu\lp x\rp f_{j}\left[\varphi(x);x\right]}=
    \int\limits_{X}\ln\int\limits_{-\varLambda}^{+\varLambda}
    \frac{d\varphi(x)}{2\varLambda}\lc 1+d\mu\lp x\rp f_{j}\left[\varphi(x);x\right]+\mathcal{O}\lb\lp d\mu\rp^{2}\rb\rc=\\
    &\int\limits_{X}\ln\lc 1+d\mu\lp x\rp
    \la f_{j}\left[\varphi(x);x\right]\ra_{\varLambda}+
    \mathcal{O}\lb\lp d\mu \rp^{2}\rb \rc
    =\int d\mu\lp x\rp\la f_{j}\left[\varphi(x);x\right] \ra_{\varLambda}.
\end{split}
\end{equation}

The resulting expression is trivial because the mean $\la f_{j}\left[\varphi(x);x\right] \ra_{\varLambda}$ does not depend on complete source $J_{t}$, therefore on initial source $j$ and auxiliary variables $t_{s}$. The complicated integral for GF $\mathcal{Z}$ is divided into independent ones, and GF $\mathcal{Z}$ is equal to one after a suitable renormalization. Such a functional is not a generating functional for any Green functions family.

However, another definitions of the expression in the last line of (\ref{GFZ-with_continuous_sum}) are possible and will be discussed in next subsections. These new definitions originate from lattice QFT and spin models type theories and their deformations \cite{kopbarsch,wipf2012statistical}.

\subsection{Derivation of GF $\mathcal{Z}$ in Basis Functions Representation: Nontrivial Integration}

In this subsection we propose different scenarios on how functional measure can be defined so it leads to nontrivial results. The central point is the definition of integrands for GF $\mathcal{Z}$ in terms of the various norms of the function. First, we first derive the expression for GF $\mathcal{Z}$ in terms of $1$-norm, and then in terms of $0$-norm.

\subsubsection{$1$-Norm and Product Integral}

Let us return to the expression in the last line of (\ref{GFZ-with_continuous_sum}) and take as \emph{the definition} the following formal transformations (we use the standard definition of the function norm):
\begin{equation}\label{First_new_definition}
\begin{split}
    &\int\limits_{X}\ln\int\limits_{-\infty}^{+\infty}
    \frac{d\varphi(x)}{\varphi_{0}}\,
    e^{\,d\mu\lp x\rp f_{j}\left[\varphi(x);x\right]}=
    \int d\mu\lp x\rp\ln\lc \int\limits_{-\infty}^{+\infty}
    \frac{d\varphi(x)}{\varphi_{0}}\,
    e^{\,d\mu\lp x\rp f_{j}\left[\varphi(x);x\right]}
    \rc^{\frac{1}{d\mu\lp x\rp}}\\
    &=\int d\mu\lp x\rp\ln\,\norm{e^{\,f_{j}\left[\varphi(x);x\right]}}_{d\mu\lp x\rp}=\int d\mu\lp x\rp\ln \lim_{p\rightarrow 0}\,\norm{e^{\,f_{j}\left[\varphi(x);x\right]}}_{d\mu\lp x\rp+p}\\
    &\stackrel{!}{=}
    \lim_{p\rightarrow 0}\int d\mu\lp x\rp
    \ln\,\norm{e^{\,f_{j}
    \left[\varphi(x);x\right]}}_{d\mu\lp x\rp+p}=
    \lim_{p\rightarrow 0}\int d\mu\lp x\rp
    \ln\,\norm{e^{\,f_{j}
    \left[\varphi(x);x\right]}}_{p}\\
    &=\lim_{p\rightarrow 0}\int 
    \frac{d\mu\lp x\rp}{p}
    \ln\int\limits_{-\infty}^{+\infty} 
    \frac{d\varphi(x)}{\varphi_{0}}\, e^{\,p\left\{-g(x)U\lb\frac{\varphi(x)}
    {\varphi_{0}}\rb + J_{t}(x)\varphi(x)\right\}}.
\end{split}
\end{equation}

The exclamation point in the third line of the expression (\ref{First_new_definition}) marks the equality in which we change the order of integration over $d\mu$ and $p$ limit calculation. This rearrangement of operations plays an important role: Since now the integration over $d\mu$ occurs at a fixed value of $p$, the norm $\norm{e^{\,f_{j}}}_{d\mu+p}$ can be replaced by the norm $\norm{e^{\,f_{j}}}_{p}$, which is done in the second equality in the third line of (\ref{First_new_definition}). Thus, the rearrangement of operations marked with an exclamation point allows to separate the measure of integration $d\mu$ and integrand, making the latter a well-defined object of mathematical analysis.

Further, one can use the following renormalization procedure (fine-tuning) for $d\mu$, $g$ and $\vec{D}$ in the last line of (\ref{First_new_definition}): $d\mu\lp x\rp=pd\mu_{\mathrm{R}}\lp x\rp$, $p g\lp x\rp=g_{\mathrm{R}}\lp x\rp$ and $p\vec{D}\lp x\rp=\vec{D}_{\mathrm{R}}\lp x\rp$. This practice is standard, for example, in the renormalization group method. Now suppose that renormalized quantities $d\mu_{\mathrm{R}}$, $g_{\mathrm{R}}$ and $\vec{D}_{\mathrm{R}}$ are independent of $p$. In this case, calculating the limit over $p$ and omitting the notation $\mathrm{R}$ (redefining the quantities):
\begin{equation}
\begin{split}
    & \int\limits_{X}\ln\int\limits_{-\infty}^{+\infty}
    \frac{d\varphi(x)}{\varphi_{0}}\,
    e^{\,d\mu\lp x\rp f_{j}\left[\varphi(x);x\right]}= 
    \int d\mu\lp x\rp\ln\int\limits_{-\infty}^{+\infty}
    \frac{d\varphi(x)}{\varphi_{0}}\,
    e^{\,f_{j}\left[\varphi(x);x\right]}\\
    &=\int d\mu\lp x\rp\ln\,\norm{e^{\,f_{j}\left[\varphi(x);x\right]}}_{1}, \quad p>0, 
\end{split}
\end{equation}
we arrive with the expression in terms of well-defined $1$-norm. Let us note that the renormalization of the coupling constant $g\rightarrow g_{\mathrm{R}}$ doesn't lead to a contradiction when considering the expansion in $1/\sqrt{g_{\mathrm{R}}}$. Equality $p g\lp x\rp=g_{\mathrm{R}}\lp x\rp$ shows that the renormalized coupling constant $g_{\mathrm{R}}$ can take various, including infinitely large, values. In the latter case, the bare coupling constant $g$ should run to infinity faster than $1/p$.

This example leads to the following conclusion. Instead of original definitions, which lead in particular to the second line of the expression (\ref{GFZ-with_continuous_sum}), consider another version of continuous product (product integral), given not only over a set $X$, but also over a spatial measure $d\mu(x)$: 
\begin{equation}\label{Something_for_Reviewer}
    \int \mathcal{D}[\varphi]\equiv 
    \prodint^{\bind d\mu(x)}\!\!\int
    \limits_{-\infty}^{+\infty} \frac{d\varphi(x)}{\varphi_{0}},\quad 
    \prodint^{\bind d\mu(x)}\!\!\mathrm{Smth}(x)\equiv
    e^{\int d\mu(x)\ln{\left[\mathrm{Smth}(x)\right]}},
\end{equation}
where $\mathrm{Smth}$ (Something) means an arbitrary, in the general case, complex-valued (with integrable phase) function of $x$

Keeping it in mind we come up with a following expressions for GF $\mathcal{Z}$:
\begin{equation}
\begin{split}
     \mathcal{Z}[j,g;d\mu]&=\int d\sigma_{t}\int \mathcal{D}[\varphi]\prodint^{\bind d\mu(x)}\!\!
     e^{\,f_{j}\left[\varphi(x);x\right]}\\ 
     &=\int d\sigma_{t}\prodint^{\bind d\mu(x)}\!\!\int\limits_{-\infty}^{+\infty} \frac{d\varphi(x)}{\varphi_{0}}\,
     e^{\,f_{j}\left[\varphi(x);x\right]}\\
     &=\int d\sigma_{t}\exp{\int d\mu\lp x\rp
     \ln\,\norm{e^{\,f_{j}
     \left[\varphi(x);x\right]}}_{1}}.
\end{split}
\end{equation}

Similar expressions are common in lattice QFT and spin models type theories and their deformations \cite{kopbarsch,wipf2012statistical} since $1$-norm is natural for the latter. However, it should be remembered that such a norm is defined in the space of functions, and, therefore, is determined by the class of the given space. Other definitions of the norm generated in the problem of GF $\mathcal{Z}$ calculation are possible. Next, we move on to one of possible definitions. We note that an independent interesting problem is the calculation of GF $\mathcal{Z}$ in the Sobolev space.

\subsubsection{$0$-Norm and Replica-Functional Taylor Series}

Consider another scenario for the expression in the last line of (\ref{GFZ-with_continuous_sum}). Take as \emph{the definition} other formal transformations (we use the standard definition of the function norm again):
\begin{equation}\label{Second_new_definition}
\begin{split}
    &\int\limits_{X}\ln\int\limits_{-\infty}^{+\infty}
    \frac{d\varphi(x)}{\varphi_{0}}\,
    e^{\,d\mu\lp x\rp f_{j}\left[\varphi(x);x\right]}
    \stackrel{!}{=}
    \lim_{p\rightarrow 0}\int d\mu\lp x\rp
    \ln\,\norm{e^{\,f_{j}
    \left[\varphi(x);x\right]}}_{p}\\
    &=\int d\mu\lp x\rp\ln \lim_{p\rightarrow 0}\,\norm{e^{\,f_{j}\left[\varphi(x);x\right]}}_{p}=
    \int d\mu\lp x\rp\ln\,\norm{e^{\,f_{j}
    \left[\varphi(x);x\right]}}_{0}.
\end{split}
\end{equation}

The exclamation point in the first line of the expression (\ref{Second_new_definition}) marks the equality in which we change the order of integration over $d\mu$ and $p$ limit calculation, with subsequent replacement of norms $\norm{e^{\,f_{j}}}_{d\mu+p}\rightarrow\norm{e^{\,f_{j}}}_{p}$. Thus, the first line of the expression (\ref{Second_new_definition}) repeats the third line of the expression (\ref{First_new_definition}). However, instead of the renormalization procedure performed in the last line of (\ref{First_new_definition}), we will use a different strategy: In the second line of (\ref{Second_new_definition}) we change back the integration and limit calculation and then calculate the limit with new integrand. In this scenario GF $\mathcal{Z}$ is expressed in terms of $0$-norm:
\begin{equation}\label{GFZ_and_second_new_definition}
    \mathcal{Z}[j,g;d\mu]= 
    \int d\sigma_{t}\exp{\int 
    d\mu\lp x\rp\ln\,\norm{e^{\,f_{j}
    \left[\varphi(x);x\right]}}_{0}},
\end{equation}
but how one can define $0$-norm itself? It is well known from the functional and infinite-dimensional analysis, that the standard axiomatic definition of the latter does not exist.

In order to solve the problem of functional integral calculation, define $0$-norm $\,\norm{f}_{0}$ of the function $f$ as follows: First, consider $0$-metric defined on two functions $f$ and $g$. Examples of such a $0$-metric (one sharp and two smooth -- the terminology is adopted from the FRG method \cite{kopbarsch,wipf2012statistical}, where both sharp and smooth FRG flow regulators are used) are:
\begin{equation}
\begin{split}
    &d\left[\,f,g\right]=\int\limits_{0}^{1}
    dz\min\lp \left|f\lp z\rp - g\lp z\rp\right|,1\rp;\\
    &d\left[\,f,g\right]=\int\limits_{0}^{1}dz\arctan\lp \left|f\lp z\rp - g\lp z\rp\right|\rp,\quad
    \,\norm{f}_{\tan}\equiv\tan d\left[\,f,0\right];\\
    &d\left[\,f,g\right]=\int\limits_{0}^{1}dz\tanh\lp \left|f\lp z\rp - g\lp z\rp\right|\rp,\quad
    \,\norm{f}_{\tanh}\equiv\text{arctanh}\,
    d\left[\,f,0\right].
\end{split}
\end{equation}

More complicated examples of $0$-metrics are built from previous examples:
\begin{equation}
    d\left[\,f,g\right]= \sum\limits_{n=1}^{\infty}\frac{1}{2^{n}}
    \int\limits_{0}^{1}dz\arctan\frac{\left|f\lp z\rp-g\lp z\rp\right|}{n},\quad
    d\left[\,f,g\right]= \sum\limits_{n=1}^{\infty}\frac{1}{2^{n}}
    \int\limits_{0}^{1}dz\tanh\frac{\left|f\lp z\rp-g\lp z\rp\right|}{n}.
\end{equation}

All the above expressions for $0$-metrics are special cases of the functional Taylor series, specified through its coefficient functions $\varPsi^{\lp n\rp}\lp z_{1},\dots,z_n\rp$:
\begin{equation}\label{Functional_Taylor_series}
    h\left(z\right)=f\left(z\right)-g\left(z\right):
    \quad d\left[\,f,g\right]=
    \sum\limits_{n=0}^{\infty}\frac{1}{n!}
    \int\limits_{0}^{1}dz_{1}\ldots
    \int\limits_{0}^{1}dz_{n}
    \varPsi^{\lp n\rp}\lp z_{1},\dots,z_n\rp 
    h\lp z_{1}\rp\dots h\lp z_{n}\rp.
\end{equation}

The guiding principle to choose the coefficient functions $\varPsi^{\lp n\rp}\lp z_{1},\dots,z_n\rp$ is the choice of space in which QFT is considered. In the case of $L^{0}$ space (zero limit of $L^{p}$ space), the functions $\varPsi^{\lp n\rp}\lp z_{1},\dots,z_n\rp$ are proportional to the Dirac delta-function. These $0$-metrics are single integrals like the examples above. If a space with $0$-metric is obtained from a space whose metric contains derivatives, the functions $\varPsi^{\lp n\rp}\lp z_{1},\dots,z_n\rp$ are proportional to derivatives of the Dirac delta-function, etc.

Recall that the original GF $\mathcal{Z}$ (\ref{generation_function_interaction}) can be represented as a ``double'' integral (\ref{Interaction_exponent}): over HS and all possible field configurations of the primary field $\varphi$, therefore over some additional space. In this approach defining QFT means defining this additional space.

The simplest idea in this way is the identification of $0$-norm of a function $f$ and $0$-metric defined on function $f$ and the identically equal to zero function (the distance between function $f$ and the origin of coordinates in the space of functions). Use this idea to define $0$-norm in expressions (\ref{Second_new_definition}) and (\ref{GFZ_and_second_new_definition}):
\begin{equation}
    \,\norm{e^{\,f_{j}\left(\varphi;x\right)}}_{0}=
    \sum\limits_{n=0}^{\infty}\frac{1}{n!}
    \int\limits_{-\infty}^{+\infty}d\varphi_{1}\ldots
    \int\limits_{-\infty}^{+\infty}d\varphi_{n}
    \varPsi^{\lp n\rp}\lp \varphi_{1},\dots,\varphi_n\rp e^{\,f_{j}\left(\varphi_{1};x\right)}\dots e^{\,f_{j}\left(\varphi_{n};x\right)}.
\end{equation}

As an example of the coefficient functions $\varPsi^{\lp n\rp}\lp \varphi_{1},\dots,\varphi_n\rp$, considering ``local type'' functions $\varPsi^{\lp n\rp}\lp \varphi_{1},\dots,\varphi_n  \rp = \alpha_{n}\int d \varphi\;\delta\lp \varphi - \varphi_{1} \rp\dots\delta\lp \varphi - \varphi_{n} \rp$, where $\alpha_{n}$ are constants, one can obtain:
\begin{equation}
    \,\norm{e^{\,f_{j}\left(\varphi;x\right)}}_{0}=
    \sum\limits_{n=0}^{\infty}\frac{\alpha_{n}}{n!}
    \int\limits_{-\infty}^{+\infty}d\varphi\, e^{\,nf_{j}\left(\varphi;x\right)}.
\end{equation}

This example leads us to $0$-norm and $0$-metric of the $L^{0}$ space. We note that the coefficients $\alpha_{n}$ define quite arbitrary function, which satisfies only general requirements: This function is bounded continuous concave and non-decreasing on the required interval.  Now considering another example of the coefficient functions $\varPsi^{\lp n\rp}\lp \varphi_{1},\dots,\varphi_n  \rp = \alpha_{n}$ yields:
\begin{equation}
    \,\norm{e^{\,f_{j}\left(\varphi;x\right)}}_{0}=
    \sum\limits_{n=0}^{\infty}\frac{\alpha_{n}}{n!}
    \lc \int\limits_{-\infty}^{+\infty}d\varphi\, e^{\,f_{j}\left(\varphi;x\right)} \rc^{n}.
\end{equation}

However, in order to obtain $0$-norm generated by some function $\varPsi$ (generator of $0$-norm), one must perform two steps. The first step is to calculate $0$-metric generated by the same function $\varPsi$. The second step is to take the value of the inverse function $\varPsi^{-1}$ of $0$-metric. The following equality illustrates this:
\begin{equation}\label{0-norm_general}
    \int d\mu\lp x\rp\ln\,\norm{e^{\,f_{j}
    \left[\varphi(x);x\right]}}_{0}=
    \int d\mu\lp x\rp\ln\varPsi^{-1}
    \int\limits_{-\infty}^{+\infty} \frac{d\varphi(x)}{\varphi_{0}}\,
    \varPsi\lc e^{\,f_{j}\left[\varphi(x);x\right]}\rc.
\end{equation}

First, consider the sharp generator $\varPsi$: By choosing $\varPsi\lp z\rp=\min\left\{z,1\right\}$ (if $z\in\mathbb{C}$ we choose $\varPsi\lp z\rp=e^{\,i\arg{z}}\min\left\{|z|,1\right\}$) in the expression (\ref{0-norm_general}), we get a sharp case. As soon as $|e^{f_{j}\lp \varphi;x\rp}|\leq1\;\Rightarrow\;\varPsi=
\varPsi^{-1}=\hat{1}$. Thus, we obtain:
\begin{equation}\label{0-norm_general_Reviewer}
    \int d\mu\lp x\rp\ln\,\norm{e^{\,f_{j}
    \left[\varphi(x);x\right]}}_{0}=
    \int d\mu\lp x\rp\ln
    \int\limits_{-\infty}^{+\infty} \frac{d\varphi(x)}{\varphi_{0}}\,
    e^{\,f_{j}\left[\varphi(x);x\right]}
    \equiv\int d\mu\lp x\rp\ln\,\norm{e^{\,f_{j}
    \left[\varphi(x);x\right]}}_{1}.
\end{equation}

Thus, in the case of the sharp generator $\varPsi$, we arrive to the lattice QFT and spin models as well, since both norms ($0$-norm and $1$-norm) coincide. We note that this is nontrivial, because it is necessary that the modulus of the generator argument does not exceed one. Otherwise, the equality (\ref{0-norm_general_Reviewer}) would be wrong: The results obtained in $1$-norm$+$renormalization procedure approach and the results obtained in $0$-norm$+$sharp $\varPsi$ approach would not coincide.

In the following part of the paper we are using two different ``smoothed'' $0$-norms:
\begin{equation}\label{0-norm_tan_tanh}
\begin{split}
    \,\norm{e^{\,f_{j}\left(\varphi;x\right)}}_{\tan} &= \tan\ \int\limits_{-\infty}^{+\infty}
    \frac{d\varphi}{\varphi_{0}}\,\arctan\lb e^{\,f_{j}\left(\varphi;x\right)}\rb\\
    \,\norm{e^{\,f_{j}\left(\varphi;x\right)}}_{\tanh} &= \text{arctanh} \int\limits_{-\infty}^{+\infty}
    \frac{d\varphi}{\varphi_{0}}\,\tanh\lb e^{\,f_{j}\left(\varphi;x\right)}\rb.
\end{split}
\end{equation}

These examples are chosen for simplicity of further calculations, the purpose of which is to demonstrate the presented theory.

All this brings us to the replica-functional Taylor series. This mathematical object can be defined as follows: Coefficient functions depend on field replicas $\varPsi^{\lp m;n_{1},\dots,n_{m} \rp}\lp \varphi_{1}^{1},\dots, \varphi_{n_{1}}^{1};\dots; \varphi_{1}^{m},\dots, \varphi_{n_{m}}^{m}  \rp$ and indices should be separable on replica indices (upper) and local inside each replica:
\begin{equation}\label{Replica-Functional_Taylor_Series}
    \sum\limits_{m=0}^{\infty}\sum\limits_{n_{1}\dots n_{m} = 0}^{\infty}\frac{1}{m!}\frac{1}{n_{1}!\dots n_{m}!}\lc\prod\limits_{a=1}^{m}  \prod\limits_{\alpha=1}^{n_{a}} \int\limits_{-\infty}^{+\infty}d\varphi_{\alpha}^{a}\rc \varPsi^{\lp m;n_{1},\dots,n_{m} \rp}\lp \lc\lc   \varphi_{\alpha}^{a}\rc_{\alpha=1}^{n_{a}}
    \rc_{a=1}^{m}\rp e^{\,\sum\limits_{a=1}^{m}
    \sum\limits_{\alpha=1}^{n_{a}}
    f_{j}\lp\varphi_{\alpha}^{a};x\rp}.
\end{equation}

Such a series makes it possible to reproduce not only $0$-metric, but also $0$-norm of any function. $0$-norms (\ref{0-norm_tan_tanh}), as well as all similar examples, are obtained from the expression (\ref{Replica-Functional_Taylor_Series}), when the coefficient functions are separable according to replica indices. We note that this series may be the subject of a separate mathematical study, as well as its applications in various fields of physics.

Thus, the replica-functional Taylor series allows one to define the functional integral (\ref{generation_function_interaction}). This definition is ambiguous, but it doesn't lead to a contradiction. We see that even in the simplest case, such an integral depends on an arbitrary (bounded continuous concave and non-decreasing on the required interval) function -- $0$-norm generator $\varPsi$. However, the generator $\varPsi$ itself is the part of the integrand. Choosing various generators $\varPsi$, we obtain various integrands, therefore, GFs $\mathcal{Z}$. The family of integrands gives rise to the corresponding family of the functional integrals. The natural $0$-norm is generated by sharp function $\varPsi$. This norm coincides with $1$-norm due to the expression (\ref{0-norm_general_Reviewer}). For this reason, the choice of the sharp $\varPsi$ is natural. Other theories (with smooth $\varPsi$) can be considered as deformations of the latter. In this approach (in terms of $1$-norm) the definition of the functional integral (\ref{generation_function_interaction}) is unique and nontrivial.

\subsubsection{Another Way to Derive $0$-Norm}

In this small subsection, we will give another way to derive the $0$-norm in expressions (\ref{Second_new_definition}) and (\ref{GFZ_and_second_new_definition}). Let us return to the second line of the expression (\ref{First_new_definition}), but now, instead of introducing a limit over $p$, we introduce an auxiliary integral with the Dirac delta-function. After that, smooth the delta-function and calculate the corresponding integral. Thus, the following statement is true:
\begin{equation}
    \int d\mu\lp x\rp\ln\,\norm{e^{\,f_{j}
    \left[\varphi(x);x\right]}}_{d\mu\lp x\rp}=
    \int d\mu\lp x\rp\ln\,\norm{e^{\,f_{j}
    \left[\varphi(x);x\right]}}_{0},
\end{equation}
due to
\begin{equation}\label{Another_way_0-norm_Reviewer}
\begin{split}
   &\ln\,\norm{e^{\,f_{j}
   \left[\varphi(x);x\right]}}_{d\mu\lp x\rp}
   =\int d\varLambda\,\delta
   \left[\varLambda-d\mu\left(x\right)\right]
   \ln\,\norm{e^{\,f_{j}
   \left[\varphi(x);x\right]}}_{\varLambda}\\
   &\rightarrow\underbrace{\int d\varLambda\,\delta_{\varepsilon}
   \left[\varLambda-d\mu\left(x\right)\right]
   \ln\,\norm{e^{\,f_{j}
   \left[\varphi(x);x\right]}}_{\varLambda}}_{\int d\varLambda\,\delta_{\varepsilon}
   \left(\varLambda\right)
   \ln\,\norm{e^{\,f_{j}
   \left[\varphi(x);x\right]}}_{\varLambda}}=
   \ln\,\norm{e^{\,f_{j}
    \left[\varphi(x);x\right]}}_{0}.
\end{split}
\end{equation}

Let us make a comment on the replacement $\delta_{\varepsilon}\left(\varLambda-d\mu\right)\rightarrow\delta_{\varepsilon}\left(\varLambda\right)$ in the expression (\ref{Another_way_0-norm_Reviewer}). We again rearrange the order of integration over $\varLambda$ and $\varepsilon$ limit calculation. The smoothed Dirac delta-function $\delta_{\varepsilon}\left(\varLambda-d\mu\right)$ with a fixed width $\varepsilon$ ($\varepsilon$ limit is calculated last) can be expanded into the Taylor series over $d\mu$. As in the expression (\ref{Trivial_integration_for_Reviewer}), all the integrals containing $\mathcal{O}[(d\mu)^{2}]$ and fixed width $\varepsilon$ vanish. For this reason, the sole survivor is the term containing $\delta_{\varepsilon}\left(\varLambda\right)$.

Also let us note that a natural generalization of the theory developed in this paper is the consideration of constructions in which any predetermined number of integrals over spatial measure $d\mu$ (embedded or nested integrals) is involved, and the internal integrand is the $0$-norm and its derivatives. In this paper, we discuss the simplest cases, leaving the rest to the future.

\subsubsection{Grand Canonical Partition Function - Like $0$-Metric}

To demonstrate the generality of the functional Taylor series, give one toy example of metric, the idea of which is derived from the statistical physics of classical gases with the interaction \cite{Ogarkov2019}:
\begin{equation}\label{Grand_Canonical_Partition_Function_Deriv}\scalebox{0.95}[0.95]{$
\begin{split}
    &\underbrace{\,\norm{e^{\,f_{j}\left(\varphi;x\right)}}_{0}}_{\text{Partition function - like 0-metric}}=
    \sum\limits_{n=0}^{\infty}\frac{1}{n!}
    \int\limits_{-\infty}^{+\infty}
    \frac{d\varphi_{1}}{\varphi_{0}}\ldots
    \int\limits_{-\infty}^{+\infty}
    \frac{d\varphi_{n}}{\varphi_{0}}
    \underbrace{\varPsi^{\lp n\rp}\lp \varphi_{1},\dots,\varphi_{n}\rp}_{e^{-\frac{1}{2}
    \sum\limits_{a=1}^{n}\sum\limits_{b=1}^{n}\varPsi\lp \varphi_{a}-\varphi_{b} \rp}} e^{\,f_{j}\left(\varphi_{1};x\right)+\ldots+
    f_{j}\left(\varphi_{n};x\right)}\\
    &=\sum\limits_{n=0}^{\infty}\frac{1}{n!}
    \int\limits_{-\infty}^{+\infty}
    \frac{d\varphi_{1}}{\varphi_{0}}\ldots
    \int\limits_{-\infty}^{+\infty}
    \frac{d\varphi_{n}}{\varphi_{0}}
    \exp\Bigg\{-\frac{1}{2}
    \sum\limits_{a=1}^{n}
    \sum\limits_{b=1}^{n}
    \underbrace{\varPsi\lp\varphi_{a}-
    \varphi_{b}\rp}_{\sum\limits_{\sigma}
    \varPsi_{\sigma}\lp\varphi_{a}\rp
    \varPsi_{\sigma}\lp\varphi_{b}\rp}+
    \sum\limits_{a=1}^{n}f_{j}
    \left(\varphi_{a};x\right)\Bigg\}\\
    &=\sum\limits_{n=0}^{\infty}\frac{1}{n!}
    \int\limits_{-\infty}^{+\infty}
    \frac{d\varphi_{1}}{\varphi_{0}}\ldots
    \int\limits_{-\infty}^{+\infty}
    \frac{d\varphi_{n}}{\varphi_{0}}
    \exp\lc -\frac{1}{2}\sum\limits_{\sigma}\lb
    \sum\limits_{a=1}^{n}
    \varPsi_{\sigma}\lp\varphi_{a}\rp\rb\lb \sum\limits_{b=1}^{n}
    \varPsi_{\sigma} \lp \varphi_{b}\rp\rb+ \sum\limits_{a=1}^{n}
    f_{j}\left(\varphi_{a};x\right)\rc\\
    &=\sum\limits_{n=0}^{\infty}\frac{1}{n!}\lc \prod\limits_{a=1}^{n}\int\limits_{-\infty}^{+\infty}
    \frac{d\varphi_{a}}{\varphi_{0}}\,
    e^{\,f_{j}\left(\varphi_{a};x\right)}\rc
    \underbrace{\lc \prod\limits_{\sigma} \int\limits_{-\infty}^{+\infty} \frac{d u_{\sigma}}{\sqrt{2\pi}}\,
    e^{-\frac{1}{2}u_{\sigma}^{2}}\rc}_{\int d\varrho_{u}} e^{\,\sum\limits_{\sigma} i u_{\sigma} \sum\limits_{a=1}^{n}\varPsi_{\sigma}
    \lp\varphi_{a}\rp}\\
    &=\sum\limits_{n=0}^{\infty}\frac{1}{n!}
    \int d\varrho_{u}\lc\prod\limits_{a=1}^{n} \int\limits_{-\infty}^{+\infty}
    \frac{d\varphi_{a}}{\varphi_{0}} \rc\underbrace{e^{\,\sum\limits_{a=1}^{n}
    f_{j}\left(\varphi_{a};x\right)+\sum\limits_{a=1}^{n}
    i\sum\limits_{\sigma}u_{\sigma}\varPsi_{\sigma}
    \lp\varphi_{a}\rp}}_{e^{\,\sum\limits_{a=1}^{n}
    \lb f_{j}\left(\varphi_{a};x\right)+
    i\vec{u}\cdot
    \vec{\varPsi}\lp\varphi_{a}\rp\rb}}\\
    &=\sum\limits_{n=0}^{\infty}\frac{1}{n!}
    \int d\varrho_{u}\prod\limits_{a=1}^{n} \int\limits_{-\infty}^{+\infty}
    \frac{d\varphi_{a}}{\varphi_{0}}\,
    e^{\,f_{j}\left(\varphi_{a};x\right)+
    i\vec{u}\cdot
    \vec{\varPsi}\lp\varphi_{a}\rp}\\
    &=\int d\varrho_{u}\exp{\int
    \limits_{-\infty}^{+\infty}
    \frac{d\varphi\left(x\right)}{\varphi_{0}}\,
    e^{-g(x)U\lb\frac{\varphi(x)}{\varphi_{0}}\rb+ J_{t}(x)\varphi(x)+i\vec{u}\cdot
    \vec{\varPsi}\left[\varphi\left(x\right)\right]}}.
\end{split}$}    
\end{equation}

All the additional notation are defined in order of detailed derivation (\ref{Grand_Canonical_Partition_Function_Deriv}). However, the results deserve a discussion. The grand canonical statistical sum can be represented in terms of some infinite-dimensional integral over measure $d\varrho_{u}$. Thus, in terms of a separate integral (besides the already existing infinite-dimensional integral over measure $d\sigma_{t}$), $0$-metric of the theory can be represented. If we identify $0$-metric and $0$-norm, the expression for GF $\mathcal{Z}$ reads:
\begin{equation}\label{Grand_Canonical_Partition_Function_Model}
    \mathcal{Z}[j,g;d\mu]=
    \int d\sigma_{t}\exp\int d\mu\lp x\rp
    \ln\int d\varrho_{u}\exp{\int
    \limits_{-\infty}^{+\infty}
    \frac{d\varphi\left(x\right)}{\varphi_{0}}\,
    e^{-g(x)U\lb\frac{\varphi(x)}{\varphi_{0}}\rb+ J_{t}(x)\varphi(x)+i\vec{u}\cdot
    \vec{\varPsi}\left[\varphi\left(x\right)\right]}}.
\end{equation}

From the expression (\ref{Grand_Canonical_Partition_Function_Model}), it is clear that the mathematical technique developed in the paper makes it possible to study sufficiently general and exotic QFTs. This can be very useful for effective field theories and models in different areas of physics, for example, in the physics of atomic nuclei and elementary particles. 

This concludes the presentation of the general theory and proceeds to the application of the latter using the polynomial QFTs in the next section.

\section{Polynomial Theory \boldmath{$\varphi_{D}^{4}$}}

In this section we will get to more specific example and obtain GF $\mathcal{Z}$ for the polynomial $\varphi^{4}_{D}$ theory. Even better, consideration with any even degree is essentially the same. For this reason we choose interaction Lagrangian to be of the form $U\left(\varphi\right) =\varphi^{2n},\, n=2,3,4,\ldots$ Derivation of GF $\mathcal{Z}$ will be shown within a framework of $1$-norm and $0$-norm, i.e. with different functional measure definitions as proposed in general theory section. As a result we will obtain  GFs $\mathcal{Z}$ as series over $1/\sqrt{g}$.

\subsection{GF $\mathcal{Z}$ of the $\varphi^{4}_{D}$ Theory in Basis Functions Representation: $1$-Norm}

First, we begin with a $1$-norm case. To be more specific, the following additional notation will be used for further derivation:
\begin{equation}\label{Additional_notation}
\begin{split}
    &\,\norm{e^{\,f_{j}\left[\varphi(x);x\right]}}_{1}=
    \int\limits_{-\infty}^{+\infty}
    \frac{d\varphi\left(x\right)}{\varphi_{0}}\, e^{-g(x)\frac{\varphi^{2n}\left(x\right)}
    {\varphi_{0}^{2n}}+
    J_{t}(x)\varphi\left(x\right)}=
    \frac{1}{\sqrt[2n]{g(x)}}
    \int\limits_{-\infty}^{+\infty}d\xi\left(x\right)
    e^{-\xi^{2n}\left(x\right)+\frac{J_{t}(x)\varphi_{0}}
    {\sqrt[2n]{g(x)}}\xi\left(x\right)},\\
    &\,\norm{e^{\,f_{j}
    \left[\varphi(x);x\right]}}_{1}^{\left(2n\right)}
    \equiv \int\limits_{-\infty}^{+\infty}
    d\xi\left(x\right)e^{-\xi^{2n}\left(x\right)+
    \frac{J_{t}(x)\varphi_{0}}
    {\sqrt[2n]{g(x)}}\xi\left(x\right)}.
\end{split}
\end{equation}

Just for a convenience here a change of variables $\frac{\varphi}{\varphi_{0}} = \frac{\xi}{\sqrt[2n]{g}}$ was made. To simplify even more let a source equals zero $j = 0$ and obtain the following integral instead of the second line in (\ref{Additional_notation}) with one more additional notation $f_{0}\left(\varphi;x\right)\equiv f\left(\varphi;x\right)$:
\begin{equation}
    \,\norm{e^{\,f
    \left[\varphi(x);x\right]}}_{1}^{\left(2n\right)}
    \equiv \int\limits_{-\infty}^{+\infty}
    d\xi\left(x\right)e^{-\xi^{2n}\left(x\right)+
    i\frac{\vec{t}\cdot\vec{D}(x)\varphi_{0}}
    {\sqrt[2n]{g(x)}}\xi\left(x\right)},
\end{equation}

\begin{figure}[H]
    \centering
    \includegraphics[scale=0.3]{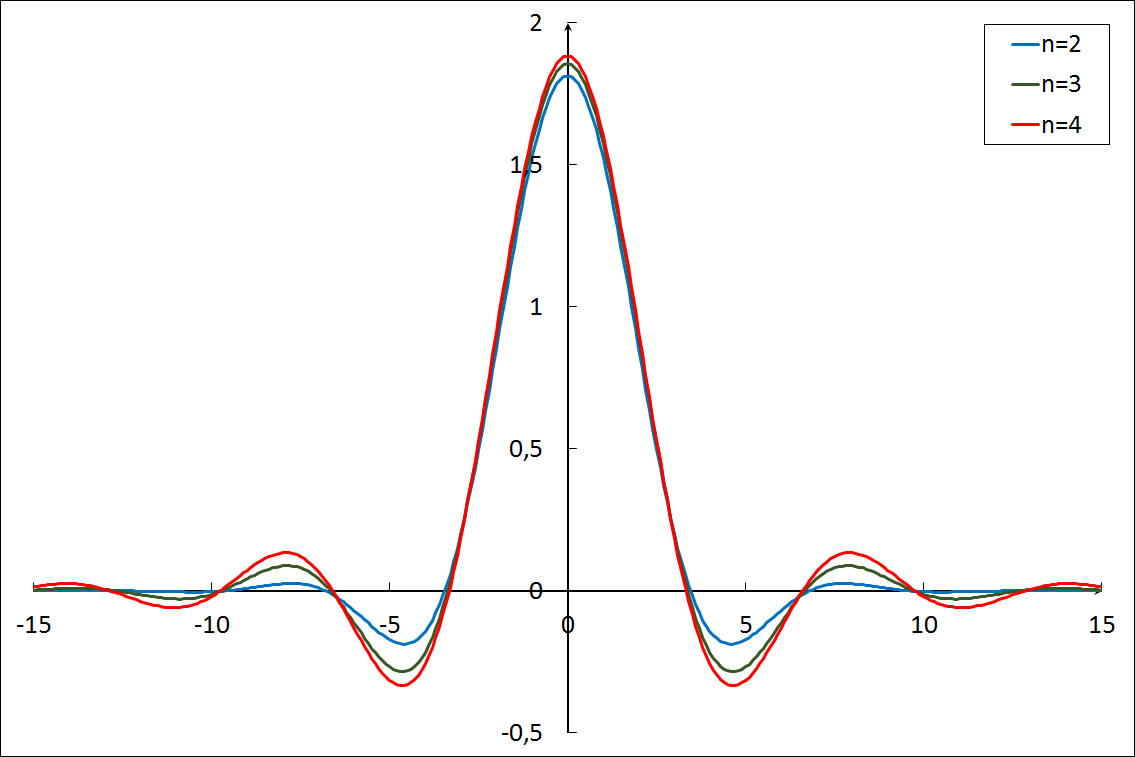}   
    \caption{A plot demonstrates how the function $\int_{-\infty}^{+\infty} d\xi\, e^{-\xi^{2n}+iw\xi}$ ($n=2,3,4$) evolves with a change of value $w$. The plotted function is not positive-definite.}
    \label{eter_polynomial}
\end{figure}

By its structure it is a Fourier transform of the function $e^{-\xi^{2n}}$. If we look on how the 
function $\int_{-\infty}^{+\infty} d\xi\, e^{-\xi^{2n}+iw\xi}$ evolves with a change of $w$, as demonstrated in Figure \ref{eter_polynomial}, its properties are easy to grasp, namely this function is even and alternating. To compare $\varphi^{4}$, $\varphi^{6}$ and $\varphi^{8}$ theories are presented.  

As one can see from the second equality in (\ref{Something_for_Reviewer}), if the function is not positive-definite, it must have an integrable phase. Thus, the question arises about the integration of this function's phase. For this reason, a strong coupling expansion is natural for the primary analysis of such a theory. This expansion is determined in the neighborhood of the point $\xi=0$, where the given function is positive-definite. However, even such a favorable expansion can be an asymptotic series. The summation of the latter is a separate task.

Next, consider GF $\mathcal{Z}$ in case when $j=0$ as we set it earlier (in terms of the product integral):
\begin{equation}
    \mathcal{Z}\lb j=0,g;d\mu\rb\equiv\mathcal{Z}
    \lb g;d\mu\rb = \mathcal{C}\lb g;d\mu\rb 
    \int d\sigma_{t}  \prodint^{\bind d\mu\lp x\rp}\!\!\,\norm{e^{\,f
    \left[\varphi(x);x\right]}}_{1}^{\left(2n\right)},
\end{equation}
where the functional $\mathcal{C}\lb g;d\mu\rb =  \prodint^{\bind d\mu\lp x\rp}\!\!\!\!\frac{1}{\sqrt[2n]{g(x)}}$ does not depend on $t_{s}$. For illustrative purpose we assume coupling function $g(x)$ to be a constant, i.e. $g=\text{const}$. In this case the functional $\mathcal{C}$ becomes a function of $g$:
\begin{equation}
    \mathcal{C}\lb g;d\mu\rb = C\left(g\right) =  \prodint^{\bind d\mu\lp x\rp}\!\!\frac{1}{\sqrt[2n]{g}} = e^{\int d\mu(x)\ln{\frac{1}{\sqrt[2n]{g}}}}=\frac{1}
    {\left(\!\!\sqrt[2n]{g}\right)^{\mathcal{V}}},
\end{equation}
the latter equality is true due to normalization condition $\int d\mu = \mathcal{V}$.

GF $\mathcal{Z}$ also becomes a function of $g$, the explicit expression for this function reads:
\begin{equation}\label{GF_function_of_g}
\begin{split}
    &\mathcal{Z}\lb d\mu\rb (g)=\frac{1}
    {\left(\!\!\sqrt[2n]{g}\right)^{\mathcal{V}}}
    \int d\sigma_{t} \prodint^{\bind d\mu\lp x\rp}\!\!\,\norm{e^{\,f
    \left[\varphi(x);x\right]}}_{1}^{\left(2n\right)},\\
    &\prodint^{\bind d\mu\lp x\rp}\!\!\,\norm{e^{\,f
    \left[\varphi(x);x\right]}}_{1}^{\left(2n\right)}= 
    \exp\int d\mu(x)\ln{\,\norm{e^{\,f
    \left[\varphi(x);x\right]}}_{1}^{\left(2n\right)}}.
\end{split}
\end{equation}

Logarithm of $1$-norm can be expanded into series over $1/\sqrt{g}$ (strong coupling expansion):
\begin{equation}
    \ln{\,\norm{e^{\,f
    \left[\varphi(x);x\right]}}_{1}^{\left(2n\right)}}= \sum\limits_{k=0}^{\infty}(-1)^{1-\delta_{0,k}}\,
    \varepsilon_{2k}^{(2n)}\, \frac{\lp\varphi_{0}\,
    \vec{t}\cdot\vec{D}\left(x\right)\rp^{2k}}
    {\lp\sqrt{g}\rp^{k}}.
\end{equation}

We explicitly indicate a sign of each term, so constants $\varepsilon_{2k}^{(2n)}$ are positive for every $k$ and $n$. After integration over spatial measure $d\mu$ and leaving only the first few terms we obtain:
\begin{equation}
\begin{split}
     \varepsilon_{0}^{(2n)} - \frac{\varepsilon_{2}^{(2n)}}{\sqrt{g}} \int d\mu(x) \lp\varphi_{0}\,\vec{t}\cdot\vec{D}\left(x\right)
     \rp^{2}+\mathcal{O}\lp\frac{1}{g}\rp\\
     = \varepsilon_{0}^{(2n)} - 
     \frac{\varepsilon_{2}^{(2n)}\varphi_{0}^{2}}
     {\sqrt{g}}\sum\limits_{s_{1},s_{2}}D_{s_{1}s_{2}}
     t_{s_{1}}t_{s_{2}}+\mathcal{O}\lp\frac{1}{g}\rp,
\end{split}
\end{equation}
where $D_{s_{1}s_{2}} = \int d\mu(x) D_{s_{1}}(x)D_{s_{2}}(x)$ is the matrix in HS. 

Substituting everything back to the first line of the expression (\ref{GF_function_of_g}) for GF $\mathcal{Z}$, the expression for the latter takes a form of the strong coupling expansion:
\begin{equation}
    \mathcal{Z}\lb d\mu\rb(g)  =\frac{e^{\varepsilon_{0}^{(2n)}}}
    {\left(\!\!\sqrt[2n]{g}\right)^{\mathcal{V}}}
    \lp 1-\frac{\varepsilon_{2}^{(2n)}
    \varphi_{0}^{2}}{\sqrt{g}}
    \sum\limits_{s}D_{ss}+\mathcal{O}\lp\frac{1}{g}\rp\rp.  
\end{equation}

Integration over the Gaussian (Efimov) measure $d\sigma_{t}$ is preformed in the following way (page $192$ in \cite{wipf2012statistical}, or any other book devoted to the statistical physics methods in QFT):
\begin{equation}\label{Trace_HS}
    \int d\sigma_{t}\sum\limits_{s_{1},s_{2}}D_{s_{1}s_{2}}t_{s_{1}}t_{s_{2}} = \sum\limits_{s_{1},s_{2}}D_{s_{1}s_{2}} \underbrace{\int d\sigma_{t}t_{s_{1}}t_{s_{2}}}_{\delta_{s_{1}s_{2}}}=
    \sum\limits_{s}D_{ss}.
\end{equation}

Since we are primarily interested in the translation-invariant case, the expression (\ref{Trace_HS}) can be made even simpler due to
\begin{equation}
    \sum\limits_{s}D_{ss} =  \sum\limits_{s} \int d\mu(x) D_{s}(x)D_{s}(x) = \int d\mu(x) L(x,x)
    =\mathcal{V}L(0).  
\end{equation}

Finally, we obtain GF $\mathcal{Z}$ (as function of $g$) as a series over inverse coupling constant $g$ (hereinafter we use the notation for the ``order parameter'' $\eta=\varphi_{0}^{2}L(0)$, that defines the ``phase transition''):
\begin{equation}\label{GFZ_phi-4_1loop}
    \mathcal{Z}\lb d\mu\rb(g)= \frac{e^{\varepsilon_{0}^{(2n)}}}
    {\left(\!\!\sqrt[2n]{g}\right)^{\mathcal{V}}}
    \left\{ 1-\frac{\varepsilon_{2}^{(2n)}\eta\mathcal{V}}
    {\sqrt{g}}+\mathcal{O}\lp\frac{1}{g}\rp\right\}.
\end{equation}

Let us note that we have performed the integration over the primary field $\varphi$ and the HS in a certain order. It is worth mentioning  that the first two terms in right hand side of (\ref{GFZ_phi-4_1loop}) are independent of spatial measure $d\mu$ which is not true for remaining ones. Also note that $1/\sqrt{g}$-order term is a sign constant, therefore, the ``phase transition'' is absent.

Let us give a more detailed description of the terms ``order parameter'' and ``phase transition'' in this paper. Usually, the order parameter refers to the solution of the Ginzburg--Landau Theory (GLT) equations (mean field theory equations). GLT order parameter is an observable quantity expressed in terms of model parameters. In this paper, we call the ``order parameter'' a combination of the model parameters themselves. Therefore, the term is quoted (we will omit quotation marks below for simplicity of presentation). This definition is given by natural causes: In the present paper, the functional integral is calculated explicitly. 

In other words, instead of highlighting the classical field configuration and solving the GLT equations, we calculate the functional integral over all possible field configurations. Therefore, the concept of the observed classical field value doesn't arise in the proposed theory. The term ``phase transition'' is also quoted. The latter means a change in the sign of $1/\sqrt{g}$ correction in the strong coupling expansion for GF $\mathcal{Z}$. In the GLT a phase transition means the appearance of a new solution (bifurcation) of the GLT equations for the observed order parameter.

To show the dependence of further terms on $d\mu$ let's write down $1/g$-order term for a $\varphi^{4}$ theory. Different (numerical) $\varepsilon$-values are redefined just for expressions to be more clear (for the compactness of the following expression, temporary notation $\alpha\left(x\right)$ is introduced):
\begin{equation}
    \ln{\,\norm{e^{\,f
    \left[\varphi(x);x\right]}}_{1}}=
    \ln \frac{\varepsilon_{0}}{\sqrt[4]{g}} - \frac{\varepsilon_{2}\,\alpha^{2}\left(x\right)}
    {\sqrt{g}}-\frac{\varepsilon_{4}\,
    \alpha^{4}\left(x\right)}{g}+ \mathcal{O}\lp\frac{1}{g^{3/2}}\rp,\quad \alpha\left(x\right)\equiv
    \varphi_{0}\,\vec{t}\cdot\vec{D}\left(x\right).
\end{equation}

The refined expression for $\varphi^{4}$ theory GF $\mathcal{Z}$ (as function of $g$) has the form:
\begin{equation}
\begin{split}
    &\mathcal{Z}\lb d\mu\rb(g)=
    \lp\frac{\varepsilon_{0}}
    {\sqrt[4]{g}}\rp^{\mathcal{V}}
    \Bigg\{1-\frac{\varepsilon_{2}
    \eta\mathcal{V}}{\sqrt{g}}-
    \frac{3\varepsilon_{4}\eta^{2}\mathcal{V}}{g}+\\ &\frac{\varepsilon_{2}^{2}
    \eta^{2}\mathcal{V}^{2}}{2g}\lb 1+
    2\int \frac{d\mu(x)}{\mathcal{V}}\int \frac{d\mu(y)}{\mathcal{V}}
    \frac{L^{2}\left(x-y\right)}{L^{2}(0)}\rb+ \mathcal{O}\lp\frac{1}{g^{3/2}}\rp\Bigg\},
\end{split}
\end{equation}
due to (page $192$ in \cite{wipf2012statistical})
\begin{equation}\label{Reminder}
    \int d\sigma_{t}t_{s_{1}}t_{s_{2}}t_{s_{3}}t_{s_{4}}= \delta_{s_{1}s_{2}}\delta_{s_{3}s_{4}}+ \delta_{s_{1}s_{3}}\delta_{s_{2}s_{4}}+
    \delta_{s_{1}s_{4}}\delta_{s_{2}s_{3}}.
\end{equation}

This concludes the consideration of the lattice QFT and spin models type (the case of the $1$-norm) and we proceed to $0$-norm in the next subsection. 

\subsection{GF $\mathcal{Z}$ of the $\varphi^{4}_{D}$ Theory in Basis Functions Representation: $0$-Norm}

In this subsection we consider the $0$-norm case and obtain also GF $\mathcal{Z}$ in terms of strong coupling expansion. Recall the expression for $0$-norm (we use operator notation for the function $\varPsi^{-1}$):
\begin{equation}\label{GFZ_phi-4_0-norm_gen}
    \,\norm{e^{\,f
    \left[\varphi(x);x\right]}}_{0}=
  \varPsi^{-1}
    \int\limits_{-\infty}^{+\infty} \frac{d\varphi(x)}{\varphi_{0}}\,
    \varPsi\lc e^{\,f\left[\varphi(x);x\right]}\rc.
\end{equation}

Substituting the $\varphi^{4}$ interaction to the expression (\ref{GFZ_phi-4_0-norm_gen}) and performing a change of integration variable as we did earlier, the $0$-norm reads:
\begin{equation}\label{GFZ_phi-4_0-change_variable}
 \varPsi^{-1} \int\limits_{-\infty}^{+\infty}
    \frac{d\varphi\lp x\rp}{\varphi_{0}}\,
    \varPsi\Bigg\{ e^{-g\lp x\rp\frac{\varphi^{4}
    \lp x\rp}{\varphi_{0}^{4}}+i\vec{t}\cdot\vec{D}
    \lp x\rp\varphi\lp x\rp}\Bigg\}
    =\varPsi^{-1}
    \int\limits_{-\infty}^{+\infty}
    \frac{d\xi\left(x\right)}
    {\sqrt[4]{g\lp x\rp}}\,
    \varPsi\lc e^{-\xi^{4}\left(x\right)+
    i\frac{\alpha\left(x\right)}
    {\sqrt[4]{g\lp x\rp}}\xi\lp x\rp}\rc.
\end{equation}

Generator $\varPsi$ of $0$-norm can be chosen in many different ways. As discussed earlier in general theory section, we choose the following examples:
$$\varPsi\lp z\rp=\begin{cases}
     & \tanh\lp z\rp \quad (\RN{1}),\\
     & \arctan\lp z\rp\quad (\RN{2}).
\end{cases}$$

Both cases, (\RN{1}) and (\RN{2}), correspond to the smooth $0$-norm: In lattice QFT and spin models theories language, this corresponds to some deformation of a theory. Let us proceed with case (\RN{1}) smooth $0$-norm, i.e. hyperbolic function. An expansion of the integral in right hand side of (\ref{GFZ_phi-4_0-change_variable}) is:
\begin{equation}\label{Intermediate_series_expansion}
    \int\limits_{-\infty}^{+\infty}
    \frac{d\xi\left(x\right)}
    {\sqrt[4]{g\lp x\rp}}\,
    \tanh\lc e^{-\xi^{4}\left(x\right)+
    i\frac{\alpha\left(x\right)}
    {\sqrt[4]{g\lp x\rp}}\xi\lp x\rp}\rc=
    \frac{\varepsilon_{1}}{\sqrt[4]{g\lp x\rp}}-
    \frac{\varepsilon_{3}\,\alpha^{2}\lp x\rp}
    {\lp\sqrt[4]{g\lp x\rp}\rp^{3}}-
    \frac{\varepsilon_{5}\,\alpha^{4}\lp x\rp}
    {\lp\sqrt[4]{g\lp x\rp}\rp^{5}}+
    \mathcal{O}\left[\frac{1}{g^{7/4}}\right].
\end{equation}

The sign of each term is indicated in the expression (\ref{Intermediate_series_expansion}) as well. We will write out explicitly only $1/g$-order terms in further expressions. A logarithm of $\varPsi^{-1}$ in the case (\RN{1}) reads:
\begin{equation}\label{lnarctanh}
\begin{split}
    &\ln\text{arctanh}
    \int\limits_{-\infty}^{+\infty}
    \frac{d\xi\left(x\right)}
    {\sqrt[4]{g\lp x\rp}}\,
    \tanh\lc e^{-\xi^{4}\left(x\right)+
    i\frac{\alpha\left(x\right)}
    {\sqrt[4]{g\lp x\rp}}\xi\lp x\rp}\rc\\
    &=\ln\lp \frac{\varepsilon_{1}}{\sqrt[4]
    {g\lp x\rp}}\rp
    +\frac{\varepsilon_{3}(x)}{\sqrt{g\lp x\rp}}+
    \frac{\varepsilon_{5}(x) }{g\lp x\rp}+
    \mathcal{O}\left[\frac{1}{g^{3/2}}\right],
\end{split}
\end{equation}
where we introduced new functions for the sake of compact notation:
\begin{equation}\label{For_future_calculations_1}
\begin{split}
    &\varepsilon_{3}(x) = \frac{1}{3}
    \varepsilon_{1}^{2}-\frac{\varepsilon_{3}}
    {\varepsilon_{1}}\alpha^{2}\lp x\rp,\\
    &\varepsilon_{5}(x) =\frac{13}{90}
    \varepsilon_{1}^{4}-
    \frac{2}{3}\varepsilon_{1}\varepsilon_{3}
    \alpha^{2}\lp x\rp-\left(\frac{1}{2}
    \frac{\varepsilon_{3}^{2}}
    {\varepsilon_{1}^{2}}+\frac{\varepsilon_{5}}
    {\varepsilon_{1}}\right)\alpha^{4}\lp x\rp.
\end{split}
\end{equation}

Next, the exponent of the expression (\ref{lnarctanh}), first integrated over spatial measure $d\mu$, with the desired accuracy is (note the presence of integrals over measures $d\mu(x)$ and $d\mu(y)$ in second line):
\begin{equation}
\begin{split}
    &e^{\,\int d\mu\lp x\rp\ln\varPsi^{-1}\lc\dots\rc}=
    e^{\,\int d\mu\lp x\rp
    \ln\frac{\varepsilon_{1}}
    {\sqrt[4]{g\lp x\rp}}}
    \Bigg\{1+\int d\mu\lp x\rp\frac{\varepsilon_{3}\lp x\rp}
    {\sqrt{g\lp x\rp}}+\\
    &\int d\mu\lp x\rp
    \frac{\varepsilon_{5}\lp x\rp}{g\lp x\rp}+
    \frac{1}{2}\int d\mu\lp x\rp
    \frac{\varepsilon_{3}\lp x\rp}
    {\sqrt{g\lp x\rp}}\int d\mu\lp y\rp
    \frac{\varepsilon_{3}\lp y\rp}
    {\sqrt{g\lp y\rp}}+
    \mathcal{O}\left[\frac{1}{g^{3/2}}\right]\Bigg\}.
\end{split}
\end{equation}

Further, an integration over the Gaussian (Efimov) measure $d\sigma_{t}$ is left to obtain GF $\mathcal{Z}$. Integration over HS (auxiliary) variables $t_{s}$ is performed in the standard way (\ref{Reminder}):
\begin{equation}\label{For_future_calculations_2}
\begin{split}
    &\int d\sigma_{t}\,\varepsilon_{3}\lp x\rp=
    \frac{1}{3}\varepsilon_{1}^{2}-
    \frac{\varepsilon_{3}}{\varepsilon_{1}}\eta,\\
    &\int d\sigma_{t}\,\varepsilon_{5}\lp x\rp=\frac{13}{90}\varepsilon_{1}^{4}-
    \frac{2}{3}\varepsilon_{1}\varepsilon_{3}\eta-
    3\lp\frac{1}{2}\frac{\varepsilon_{3}^{2}}
    {\varepsilon_{1}^{2}}+\frac{\varepsilon_{5}}
    {\varepsilon_{1}}\rp\eta^{2},\\
    &\int d\sigma_{t}\,\varepsilon_{3}\lp x\rp
    \varepsilon_{3}\lp y\rp=
    \frac{1}{9}\varepsilon_{1}^{4}-
    \frac{2}{3}\varepsilon_{1}\varepsilon_{3}\eta+
    \frac{\varepsilon_{3}^{2}}{\varepsilon_{1}^{2}}
    \eta^{2}\left[1+2\frac{L^{2}\left(x-y\right)}
    {L^{2}(0)}\right].
\end{split}
\end{equation}

Thus, the expression for GF $\mathcal{Z}$ takes the form:
\begin{equation}\label{For_future_calculations_3}
\begin{split}
    &\mathcal{Z}\lb g;d\mu\rb=
    e^{\int d\mu\lp x\rp\ln\frac{\varepsilon_{1}}
    {\sqrt[4]{g\lp x\rp}}}\Bigg\{ 1+
    \lb\frac{1}{3}\varepsilon_{1}^{2}-
    \frac{\varepsilon_{3}}{\varepsilon_{1}}\eta\rb
    \int\frac{d\mu\lp x\rp}{\sqrt{g\lp x\rp}}\\
    &+\left[\frac{13}{90}\varepsilon_{1}^{4}-
    \frac{2}{3}\varepsilon_{1}\varepsilon_{3}\eta-
    3\lp\frac{1}{2}\frac{\varepsilon_{3}^{2}}
    {\varepsilon_{1}^{2}}+\frac{\varepsilon_{5}}
    {\varepsilon_{1}}\rp\eta^{2}\right]
    \int\frac{d\mu\lp x\rp}{g\lp x\rp}\\
    &+\frac{1}{2}\left[\frac{1}{9}\varepsilon_{1}^{4}-
    \frac{2}{3}\varepsilon_{1}\varepsilon_{3}\eta+
    \frac{\varepsilon_{3}^{2}}{\varepsilon_{1}^{2}}
    \eta^{2}\right]
    \int\frac{d\mu\lp x\rp}{\sqrt{g\lp x\rp}}
    \int\frac{d\mu\lp y\rp}{\sqrt{g\lp y\rp}}\\
    &+\frac{\varepsilon_{3}^{2}}{\varepsilon_{1}^{2}}
    \eta^{2}
    \int\frac{d\mu\lp x\rp}{\sqrt{g\lp x\rp}}
    \int\frac{d\mu\lp y\rp}{\sqrt{g\lp y\rp}}
    \frac{L^{2}\left(x-y\right)}{L^{2}(0)}+
    \mathcal{O}\left[\frac{1}{g^{3/2}}\right]\Bigg\}.
\end{split}
\end{equation}

If we let a coupling constant function be a constant, i.e. $g\lp x\rp\equiv g\equiv const$, then we obtain more simplified  expression for GF $\mathcal{Z}$:
\begin{equation}\label{Final_result_polynomial_theory}
\begin{split}
    &\mathcal{Z}\lb d\mu\rb\left(g\right)=
    \lp\frac{\varepsilon_{1}}
    {\sqrt[4]{g}}\rp^{\mathcal{V}}\Bigg\{ 1+
    \lb\frac{1}{3}\varepsilon_{1}^{2}-
    \frac{\varepsilon_{3}}{\varepsilon_{1}}\eta\rb
    \frac{\mathcal{V}}{\sqrt{g}}\\
    &+\left[\frac{13}{90}\varepsilon_{1}^{4}-
    \frac{2}{3}\varepsilon_{1}\varepsilon_{3}\eta-
    3\lp\frac{1}{2}\frac{\varepsilon_{3}^{2}}
    {\varepsilon_{1}^{2}}+\frac{\varepsilon_{5}}
    {\varepsilon_{1}}\rp\eta^{2}\right]
    \frac{\mathcal{V}}{g}\\
    &+\frac{1}{2}\left[\frac{1}{9}\varepsilon_{1}^{4}-
    \frac{2}{3}\varepsilon_{1}\varepsilon_{3}\eta+
    \frac{\varepsilon_{3}^{2}}{\varepsilon_{1}^{2}}
    \eta^{2}\right]
    \frac{\mathcal{V}^{2}}{g}\\
    &+\frac{\varepsilon_{3}^{2}}{\varepsilon_{1}^{2}}
    \eta^{2}\frac{\mathcal{V}^{2}}{g}
    \int\frac{d\mu\lp x\rp}{\mathcal{V}}
    \int\frac{d\mu\lp y\rp}{\mathcal{V}}
    \frac{L^{2}\left(x-y\right)}{L^{2}(0)}+
    \mathcal{O}\left(\frac{1}{g^{3/2}}\right)\Bigg\}.
\end{split}
\end{equation}

As one can see from the expression (\ref{Final_result_polynomial_theory}), already in order of $1/\sqrt{g}$ a phase transition appears in the model with a smooth $0$-norm. Let us make an important remark: The presented theory is formulated in the whole space (without boundaries). In such a space, a spatial integration measure $d\mu$ is specified, normalized with $\mathcal{V}$, where $\mathcal{V}$ is the model parameter. The latter is not the volume of the system. One more point: Hermite functions are chosen as basis functions. Such a basis is a natural basis in the HS defined over a space without boundaries. In the case of a finite volume, the basis is trigonometric functions numbered by a discrete momentum. We note that the $\mathcal{S}$-matrix of such a theory was constructed in the papers \cite{efimov1977cmp,efimov1979cmp}. 

Thus, in this paper we consider a system infinite in space, and the phase transition is understood in the above sense. This concludes the application of the general mathematical technique, developed in the present paper, to the polynomial theory $\varphi^{4}_{D}$. We proceed to nonpolynomial ones in the next section.

\section{Nonpolynomial Theory  \boldmath{$\text{Sinh}^{4}\varphi_{D}$}}

In this section we derive and analyze GF $\mathcal{Z}$ of $\sinh^{4}\varphi_{D}$ theory. The results of this section can be easily generalized to the case of theories $\sinh^{6}\varphi_{D},\,\sinh^{8}\varphi_{D}, \ldots$ It is important that the asymptotics of the interaction Lagrangian be infinite. If we consider the case of an all-bounded interaction Lagrangian, for example, the sine-Gordon theory, the methods of statistical physics or the variational principle can be used to study the problem \cite{efimov1970nonlocal,efimov1977cmp,efimov1977nonlocal,efimov1985problems,basuev1973conv,basuev1975convYuk,Ogarkov2019}.

\subsection{GF $\mathcal{Z}$ of the $\text{Sinh}^{4}\varphi_{D}$ Theory in Basis Functions Representation: $1$-Norm}

In this section we consider a nonpolynomial theory $\sinh^{4}\varphi$. In this case, the interaction Lagrangian $U\left(\varphi\right)$ is given by the expression (in our case $n=2$, but $n=2,3,4,\ldots$ generalizations are possible):
\begin{equation}
    U\lb\varphi\lp x\rp\rb =\sinh^{2n}
    \left[\frac{\varphi\lp x\rp}
    {\varphi_{0}}\right],
\end{equation}
where $\varphi_{0}$ is the dimensional divisor (a field scale). Thus, the $1$-norm is given by (recall the notation $\alpha\left(x\right)=\varphi_{0}\,\vec{t}\cdot\vec{D}\left(x\right)$; the source $j=0$ as in the polynomial case):
\begin{equation}\label{Nonpolynomial_Theory_1-Norm}
\begin{split}
    \,\norm{e^{\,f
    \left[\varphi(x);x\right]}}_{1}&= \int\limits_{-\infty}^{+\infty}
    \frac{d\varphi\left(x\right)}{\varphi_{0}}\,
    e^{-g\lp x\rp\sinh^4
    \left[\frac{\varphi\left(x\right)}
    {\varphi_{0}}\right]+i\vec{t}
    \cdot\vec{D}\lp x\rp
    \varphi\left(x\right)}\\
    &=\int\limits_{-\infty}^{+\infty}
    \frac{d\xi\left(x\right)}
    {\sqrt[4]{g\lp x\rp}}\, 
    e^{-g\lp x\rp\sinh^4\left[\frac{\xi\left(x\right)}
    {\sqrt[4]{g\lp x\rp}}\right]+
    iw\left(x\right)\xi\left(x\right)},
\end{split}
\end{equation}
where temporary notation $w\left(x\right)=\frac{\alpha\left(x\right)}{\sqrt[4]{g\lp x\rp}}$ and $\frac{\varphi\left(x\right)}{\varphi_{0}}=
\frac{\xi\left(x\right)}{\sqrt[4]{g\left(x\right)}}$. Expanding the integrand into the Taylor series over $1/\sqrt{g}$ (corresponding to a large coupling constant $g$) and omitting terms of odd $\xi$ powers (since integration of odd function in symmetric limits yields zero) for $1$-norm (\ref{Nonpolynomial_Theory_1-Norm}) we obtain:
\begin{equation}
\begin{split}
    \,\norm{e^{\,f
    \left[\varphi(x);x\right]}}_{1}&=
    \int\limits_{-\infty}^{+\infty}
    \frac{d\xi\left(x\right)}
    {\sqrt[4]{g\lp x\rp}}\, 
    e^{-\xi^4\left(x\right)}
    \left\{1-\frac{\left[w\left(x\right)
    \xi\left(x\right)\right]^2}{2!}-\frac{2}{3}
    \frac{\left[\xi\left(x\right)\right]^6}
    {\sqrt{g\lp x\rp}}+
    \mathcal{O}\left[\frac{1}{g}\right]\right\}\\
    &=\frac{2\varGamma\lp\frac{5}{4}\rp}
    {\sqrt[4]{g\lp x\rp}}\left\{1-\frac{\pi}
    {\sqrt{2g\lp x\rp}\varGamma^2\lp\frac{1}{4}\rp}
    \lb\alpha^{2}\left(x\right)-1\rb+
    \mathcal{O}\left[\frac{1}{g}\right]\right\}.
\end{split}
\end{equation}

\begin{figure}[H]
    \centering
    \includegraphics[scale=0.3]{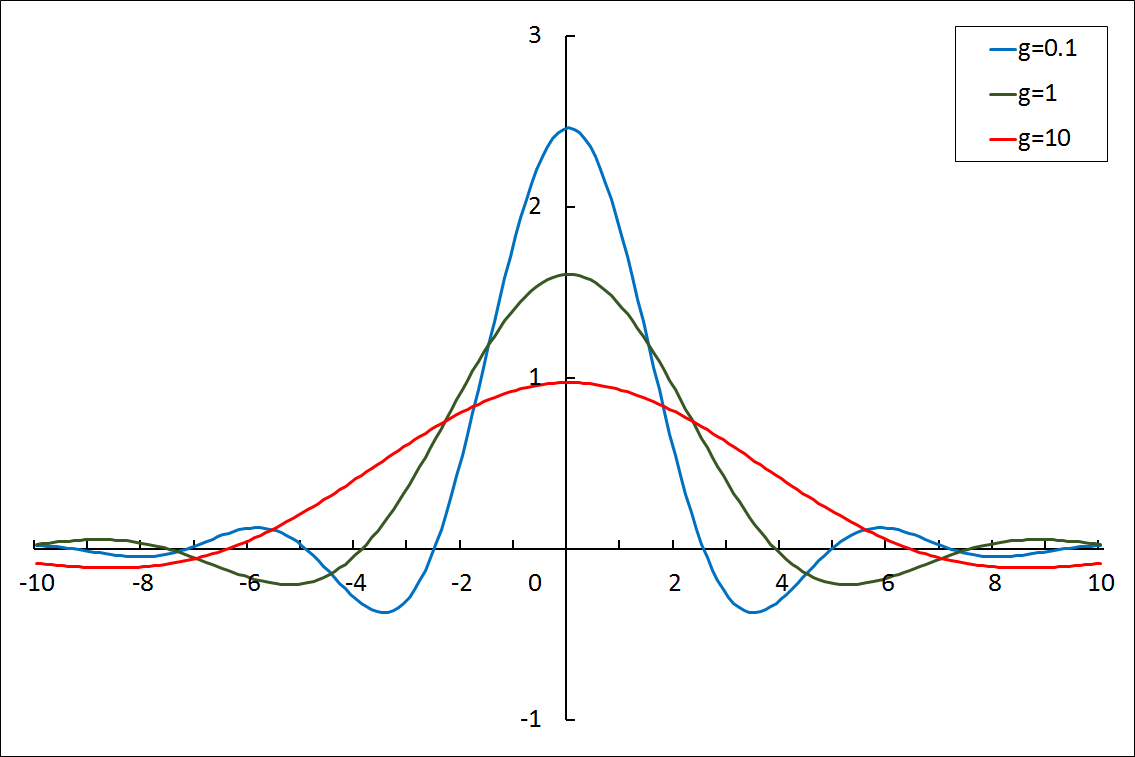}   
    \caption{The $1$-norm in the nonpolynomial $\sinh^{4}\varphi$ theory plotted with respect to $w$ for different values of the coupling constant $g$. The field scale $\varphi_{0}=1$.}
    \label{eter_nonpolynomial}
\end{figure}

Here $\varGamma(z)$ is the Euler gamma-function. Next, consider the functional integral over measure $\mathcal{D}\left[\varphi\right]$ in terms of a product integral with $1$-norm:
\begin{equation}\label{Nonpolynomial_Theory_prod_int}
\begin{split}
    &\prodint^{\bind d\mu\lp x\rp}\!\!
    \,\norm{e^{\,f\left[\varphi(x);x\right]}}_{1}=
    \mathcal{C}_{1}\lb g;d\mu\rb
    e^{\int d\mu\lp x\rp \ln{\left\{1-\frac{\pi}
    {\sqrt{2g\lp x\rp}\varGamma^2\lp\frac{1}{4}\rp}
    \lb\alpha^{2}\left(x\right)-1\rb+
    \mathcal{O}\left[\frac{1}{g}\right]\right\}}},\\
    &\mathcal{C}_{1}\lb g;d\mu\rb =
    \prodint^{\bind d\mu\lp x\rp}\!
    \frac{2\varGamma\lp\frac{5}{4}\rp}
    {\sqrt[4]{g\lp x\rp}}=
    e^{\int d\mu\lp x\rp
    \ln{\frac{2\varGamma\lp\frac{5}{4}\rp}
    {\sqrt[4]{g\lp x\rp}}}}.
    \end{split}    
\end{equation}

The functional $\mathcal{C}_{1}\lb g;d\mu\rb$ in the second line of the expression (\ref{Nonpolynomial_Theory_prod_int}) does not depend on $t_{s}$. Expanding the exponent in the first line of the expression (\ref{Nonpolynomial_Theory_prod_int}) into a power series up to $\mathcal{O}[1/g]$ the following expression reads:
\begin{equation}
    \int d\sigma_{t}\prodint^{\bind d\mu\lp x\rp}\!\!\,\norm{e^{\,f
    \left[\varphi(x);x\right]}}_{1}=
    \mathcal{C}_{1}\lb g;d\mu\rb
    \left\{1-\frac{\pi}
    {\varGamma^{2}\lp\frac{1}{4}\rp}
    \int\frac{d\mu\lp x\rp}
    {\sqrt{2g\lp x\rp}}\lb\int d\sigma_{t}\,
    \alpha^{2}\left(x\right)-1\rb+
    \mathcal{O}\left[\frac{1}{g}\right]\right\}.
\end{equation}

As soon as the measure $d\sigma_{t}$ is Gaussian, and the function $D_{s}\lp x\rp$ does not depend on $t_{s}$, the following is true:
\begin{equation}
    \int d\sigma_{t}\,\alpha^{2}\left(x\right)=
    \varphi_{0}^{2}\sum_{s_{1}, s_{2}}
    D_{s_{1}}\lp x\rp D_{s_{2}}
    \lp x\rp\int d\sigma_{t}\,t_{s_{1}} t_{s_{2}}=\varphi_{0}^{2}L(0).
\end{equation}

As before, introduce the ``order parameter'' $\eta=\varphi_{0}^{2}L\lp 0\rp$. Already in the $1$-norm case, a phase transition arise in the nonpolynomial theory. In some sense, nonpolynomiality plays the role of a smooth generator $\varPsi$. The expression for GF $\mathcal{Z}$ reads:
\begin{equation}\label{Nonpolynomial_GFZ}
    \begin{split}
    \mathcal{Z}\lb g;d\mu \rb=
    \int d\sigma_{t}\prodint^{\bind d\mu\lp x\rp}\!\!\,\norm{e^{\,f
    \left[\varphi(x);x\right]}}_{1}=
    \mathcal{C}_{1}\lb g;d\mu\rb
    \left\{1+\frac{\pi\lp 1-\eta\rp}
    {\varGamma^{2}\lp \frac{1}{4} \rp}
    \int\frac{d\mu\lp x\rp}
    {\sqrt{2g\lp x\rp}}+
    \mathcal{O}\left[\frac{1}{g}\right]\right\}.
    \end{split}
\end{equation}

From the expression (\ref{Nonpolynomial_GFZ}) it is clear that the critical value of $\eta$ (phase transition: $1/\sqrt{g}$-correction changes sign) is $\eta_{c}=1$. For the case $g\lp x\rp\equiv g\equiv const$ GF $\mathcal{Z}$ (\ref{Nonpolynomial_GFZ}) becomes a function of the coupling constant $g$:
\begin{equation}
    \begin{split}
    \mathcal{Z}\lb d\mu \rb\left(g\right)=
    \lp\frac{2\varGamma\lp\frac{5}{4}\rp}
    {\sqrt[4]{g}}\rp^{\mathcal{V}}
    \left\{1+\frac{\pi\lp 1-\eta\rp}
    {\varGamma^{2}\lp \frac{1}{4}\rp}
    \frac{\mathcal{V}}{\sqrt{2g}}+
    \mathcal{O}\left(\frac{1}{g}\right)\right\}.
    \end{split}
\end{equation}

In conclusion of this subsection, we note the following: Figure \ref{eter_nonpolynomial} demonstrates the same behavior in $\sinh^{4}\varphi$ theory as in the polynomial $\varphi^{4}_{D}$ theory. Since the plotted function is not positive-definite, a strong coupling expansion is natural for the primary analysis of such a theory. However, this should not exclude a deeper analysis, which should be the subject of the furthest.

\subsection{GF $\mathcal{Z}$ of the $\text{Sinh}^{4}\varphi_{D}$ Theory in Basis Functions Representation: $0$-Norm}

In this subsection we consider the $0$-norm case. As mentioned above, the definition of $0$-norm, which makes it possible to calculate a functional integral, is generally formulated in terms of the replica-functional Taylor series (\ref{Replica-Functional_Taylor_Series}). We will restrict ourselves to a special case when the $0$-norm definition is formulated in terms of $0$-norm generator $\varPsi\left(z\right)=\arctan\left(z\right)$. Such a generator is the ``building block'' for more general expressions, like $\varPsi_{1}\left(z\right)=\sum\limits_{n=1}^{\infty}2^{-n}\arctan\lp\frac{z}{n}\rp$, but $\varPsi_{1}^{-1}\lp z\rp$ would be technically hard to obtain analytically. With $\varPsi\lp z\rp=\arctan\lp z\rp$ for the theory $\sinh^{4}\varphi$ one can obtain:
\begin{equation}\label{Calculation_in_detail_1}
\begin{split}
    &\int\limits_{-\infty}^{+\infty}
    \frac{d\varphi\left(x\right)}
    {\varphi_{0}}\arctan\lc e^{-g\lp x\rp\sinh^{4}\left[\frac{\varphi\lp x\rp}{\varphi_{0}}\right]+
    i\vec{t}\cdot\vec{D}
    \lp x\rp\varphi\lp x\rp} \rc\\
    &=\int\limits_{-\infty}^{+\infty}
    \frac{d\xi\lp x\rp}{\sqrt[4]{g\lp x\rp}}
    \arctan\Bigg\{ e^{-g\lp x\rp\sinh^{4}
    \left[\frac{\xi\lp x\rp}{\sqrt[4]
    {g\lp x\rp}}\right]+iw\left(x\right)\xi\lp x\rp }\Bigg\},
\end{split}
\end{equation}
where the notation of the previous subsection are used. 

Consider the calculation in detail. The expansion into series over the inverse coupling constant $1/\sqrt{g}$ (strong coupling expansion) in the second line of (\ref{Calculation_in_detail_1}) is:
\begin{equation}\label{Calculation_in_detail_2}
    \int\limits_{-\infty}^{+\infty}
    \frac{d\xi\lp x\rp}{\sqrt[4]{g\lp x\rp}}
    \arctan\lc\dots\rc=
    \frac{1}{\sqrt[4]{g\lp x\rp}}
    \lc\varepsilon_{0}-\frac{\varepsilon_{1}+
    \varepsilon_{2}\alpha^{2}
    \lp x\rp}{12\sqrt{g\lp x\rp}}+
    \mathcal{O}\left[\frac{1}{g}\right]\rc,
\end{equation}
where $\varepsilon_{0}\approx1.50291$, $\varepsilon_{1}\approx3.28319$, $\varepsilon_{2}\approx1.45135$ (we again use $\varepsilon$-notation for various numerical values just for expressions to be more clear).

The expansion into series over $1/\sqrt{g}$ of the $\ln\varPsi^{-1}$ of (\ref{Calculation_in_detail_2}) reads:
\begin{equation}\label{Calculation_in_detail_3}
    \ln\tan\int\limits_{-\infty}^{+\infty}
    \frac{d\xi\lp x\rp}{\sqrt[4]{g\lp x\rp}}\arctan\lc\dots\rc=
    \ln\frac{\varepsilon_{0}}
    {\sqrt[4]{g\lp x\rp}}+
    \frac{4\varepsilon_{0}^{3}-
    \varepsilon_{1}-\varepsilon_{2}\alpha^{2}\lp x\rp}
    {12\varepsilon_{0}\sqrt{g\lp x\rp}}+
    \mathcal{O}\left[\frac{1}{g}\right].
\end{equation}

The exponent of the expression (\ref{Calculation_in_detail_3}) with the desired accuracy is:
\begin{equation}\label{Calculation_in_detail_4}
    e^{\int d\mu\lp x\rp\ln\tan\lc\dots\rc}=
    e^{\int d\mu\lp x\rp\ln
    \frac{\varepsilon_{0}}{\sqrt[4]{g\lp x\rp}}}
    \lc 1+\int d\mu\lp x\rp\frac{4\varepsilon_{0}^{3}-
    \varepsilon_{1}-\varepsilon_{2}\alpha^{2}
    \lp x\rp}{12\varepsilon_{0}
    \sqrt{g\lp x\rp}}+
    \mathcal{O}\left[\frac{1}{g}\right]\rc.
\end{equation}

The expression (\ref{Calculation_in_detail_4}) allows one to write down an expression for GF $\mathcal{Z}$ in the following way:
\begin{equation}\label{Calculation_in_detail_5}
    \mathcal{Z}\lb g;d\mu\rb=
    e^{\int d\mu\lp x\rp\ln
    \frac{\varepsilon_{0}}{\sqrt[4]{g\lp x\rp}}}
    \lc 1+\int d\mu\lp x\rp
    \frac{ 4\varepsilon_{0}^{3}-\varepsilon_{1}-
    \varepsilon_{2}\varphi_{0}^{2}L\left(0\right)}
    {12\varepsilon_{0}\sqrt{g\lp x\rp}}+
    \mathcal{O}\left[\frac{1}{g}\right]\rc.
\end{equation}

Consider again the case $g\lp x\rp\equiv g\equiv const\;\Rightarrow$ GF $\mathcal{Z}$ (\ref{Calculation_in_detail_5}) as a function of $g$ has the form:
\begin{equation}\label{Calculation_in_detail_6}
    \mathcal{Z}\lb d\mu\rb\left(g\right)=
    \left(\frac{\varepsilon_{0}}
    {\sqrt[4]{g}}\right)^{\mathcal{V}}
    \lc 1+\frac{\mathcal{V}}
    {12\varepsilon_{0}\sqrt{g}}
    \underbrace{\lb 4\varepsilon_{0}^{3}-
    \varepsilon_{1}-\varepsilon_{2}
    \varphi_{0}^{2}L\left(0\right)\rb}_{\approx
    10.2955-1.45135\eta} +\mathcal{O}\left(\frac{1}{g}\right)\rc.
\end{equation}

Calculated critical values of order parameter $\eta$ are follows: $\eta_{c}^{\tan}\approx7.0937$ for ``$\tan$-norm'' and $\eta_{c}^{\tanh}\approx8.9327$ for ``$\tanh$-norm'' (the result of a separate calculation).

Travelling back to the polynomial scalar field theory $\varphi^{4}$:
\begin{equation}\label{Calculation_in_detail_7}
    \mathcal{Z}\lb d\mu\rb\left(g\right)=
    \lp\frac{\varepsilon_{1}}
    {\sqrt[4]{g}}\rp^{\mathcal{V}}
    \lc 1+\frac{\mathcal{V}}
    {3\varepsilon_{1}\sqrt{g}}
    \underbrace{\lb\varepsilon_{1}^{3}-
    3\varepsilon_{3}\varphi_{0}^{2}
    L\left(0\right)\rb}_{\approx
    3.39466-0.362838\eta}
    +\mathcal{O}\lp\frac{1}{g}\rp\rc.
\end{equation}

Calculated critical values of order parameter $\eta$ are follows: $\eta_{c}^{\tan}\approx9.3559$ for ``$\tan$-norm'' (the expression (\ref{Calculation_in_detail_7}) is the result of a separate calculation) and $\eta_{c}^{\tanh}\approx11.9895$ for ``$\tanh$-norm'' from the first line of the expression (\ref{Final_result_polynomial_theory}). We note that polynomial $\varphi^{4}$ theory critical values are higher than nonpolynomial $\sinh^{4}\varphi$ theory ones. Also note, that ``$\tanh$-norm'' critical values appear to be higher than ``$\tan$-norm'' ones.

In conclusion of the section we make an important remark. In all considered theories, the existence of $L\left(0\right)=G^{-1}\left(0\right)$ is assumed. The inverse nonlocal propagator of a type
\begin{equation}\label{Calculation_in_detail_8}
    L\left(x\right)=\frac{1}{x^{2}}
    \left(1-e^{-\frac{x^{2}}{l^{2}}}\right)
\end{equation}
is an example of such a function. In this case, $L\lp0\rp=l^{-2}$ and $\varphi_{0}^{2}L\lp0\rp=\varphi_{0}^{2}\,l^{-2}$. As soon as $\varphi_{0}$ is an ultraviolet value, the order parameter $\eta=\varphi_{0}^{2}L\lp0\rp\sim1$. Therefore, critical values $\eta_{c}$ are accessible. However, if the propagator $G\left(x\right)$ itself has the form (\ref{Calculation_in_detail_8}), an additional regularization in the theory is necessary. At the same time, the ratio $\eta$ should be true for both initial and regularized quantities.

\section{Continuous Lattice of Functions}

The mathematical technique described in previous sections was based on integration over separable HS (the number of auxiliary variables $t_{s}$, therefore, of integrals over $t_{s}$ is countable). In this section we consider the generalization of the theory to the case when the number of auxiliary variables $t_{s}$ is uncountable. Recall that we call these cases discrete and continuous lattices of functions, respectively. 

In conclusion of this section, we derive this generalization in an independent way, starting from Parseval--Plancherel identity (page $176$ in \cite{Shavgulidze2015}). This identity is the strongest tool that allows to establish different dualities in QFTs at the functional integral level.

\subsection{Derivation of GF $\mathcal{Z}$ in Basis Functions Representation}

Consider the expansion of the inverse propagator $\hat{L}$, more precisely, its kernel $L\left(x,y\right)$, in the Fourier integral:
\begin{equation}
\begin{split}
    L\left(x,y\right)&=
    \int d\mu\left(k\right)L\left(k\right)
    e^{\,ik\left(x-y\right)}=
    \int d\mu\left(k\right)L\left(k\right)
    \cos{\left[k\left(x-y\right)\right]}\\
    &=\int d\mu\left(k\right)L\left(k\right)
    \left[\cos{\left(kx\right)}\cos{\left(ky\right)}+ \sin{\left(kx\right)}\sin{\left(ky\right)}\right]\\
    &=\sum\limits_{\sigma}
    \int d\mu\left(k\right) L\left(k\right)
    \psi_{\sigma}\left(kx\right)
    \psi_{\sigma}\left(ky\right)
    \equiv\sum\limits_{\sigma}
    \int d\mu\left(k\right)
    D_{\sigma,k}\left(x\right)
    D_{\sigma,k}\left(y\right),
\end{split}
\end{equation}
where $\sigma\in\{c,s\}$ ($\psi_{c}\equiv \cos,\,\psi_{s}\equiv \sin$), $d\mu\left(k\right)$ is the appropriate measure in $k$ (momentum) space, $L\left(k\right)$ is the Fourier transform of the function $L\left(x,y\right)$, and $D_{\sigma,k}$ is the continuous analog of the function $D_{s}$. Thus, the discrete index $s$ is now replaced by a pair $(\sigma,k)$.

The free theory action $S_{0}$ can be rewritten using basis of functions representation, similar to the expression (\ref{Inverse_Propagator_Splitting}) at the beginning of the paper:
\begin{equation}
\begin{split}
    S_{0}\lb\varphi\rb &=\frac{1}{2}
    \lp\varphi\left|\hat{L}\right|\varphi\rp=
    \frac{1}{2}\int d^{D}x \int d^{D}y\, 
    \varphi\left(x\right)\varphi\left(y\right)
    \sum\limits_{\sigma}\int d\mu\left(k\right) D_{\sigma,k}\left(x\right)
    D_{\sigma,k}\left(y\right)\\
    &=\frac{1}{2}\sum\limits_{\sigma}
    \int d\mu\left(k\right)
    \lp\int d^{D}x D_{\sigma,k}\left(x\right)
    \varphi\left(x\right)\rp
    \lp\int d^{D}y D_{\sigma,k}\left(y\right)
    \varphi\left(y\right)\rp\\ 
    &\equiv\frac{1}{2}\sum\limits_{\sigma}
    \int d\mu\left(k\right)
    \lp D_{\sigma,k}|\varphi\rp^{2}.
\end{split}
\end{equation}

Continuing to draw parallels, an exponent with a free theory action $S_{0}$ can be rewritten using Gaussian integral trick (Hubbard--Stratonovich transformation) with auxiliary variables $t_{\sigma,k}$,  similar to the expression (\ref{Hubbard_Stratonovich_transformation}):
\begin{equation}
\begin{split}
    &e^{-S_{0}\lb\varphi\rb}=
    \prod\limits_{\sigma,k}
    e^{-\frac{1}{2}d\mu\left(k\right)
    \lp D_{\sigma,k}|\varphi\rp^{2}}=
    \prod\limits_{\sigma,k}
    \int\limits_{-\infty}^{+\infty}
    \frac{dt_{\sigma,k}}
    {\sqrt{2\pi d\mu\left(k\right)}}\, e^{-\frac{1}{2d\mu\left(k\right)}t_{\sigma,k}^{2}+ it_{\sigma,k}\lp D_{\sigma,k}|\varphi\rp}\\
    &=\prod\limits_{\sigma,k}
    \int\limits_{-\infty}^{+\infty}
    dt_{\sigma,k}\sqrt{\frac{d\mu\left(k\right)}{2\pi}} e^{\,d\mu\left(k\right)
    \left[-\frac{1}{2}t_{\sigma,k}^{2}+
    it_{\sigma,k}\lp D_{\sigma,k}|\varphi\rp\right]}
    \equiv\int\mathcal{D}\sigma\lb t\rb\, e^{\,i\sum\limits_{\sigma}\int d\mu\left(k\right)t_{\sigma,k}\lp D_{\sigma,k}|\varphi\rp}.
\end{split}
\end{equation}

Next, we introduce the complete source: $J_{t}\left(x\right)=j\left(x\right)+i \sum\limits_{\sigma}\int d\mu\left(k\right) t_{\sigma,k}D_{\sigma,k}\left(x\right)$. In terms of this definition GF $\mathcal{Z}$ reads:
\begin{equation}
\begin{split}
    &\mathcal{Z}\left[j,g\right]=
    \int\mathcal{D}\sigma\lb t\rb
    \int\mathcal{D}\left[\varphi\right]\, 
    e^{-S_{1}\lb g,\varphi\rb + 
    \left(J_{t}|\varphi\right)}\\
    &=\int\mathcal{D}\sigma\lb t\rb
    \prod\limits_{X}\exp{\ln\int
    \limits_{-\infty}^{+\infty}
    \frac{d\varphi\left(x\right)}{\varphi_{0}}\,
    e^{\,d^{D}x\lp x\rp f_{j}
    \left[\varphi\left(x\right);x\right]}}\\
    &=\int\mathcal{D}\sigma\lb t\rb
    \exp{\int\limits_{X}
    \ln\int\limits_{-\infty}^{+\infty}
    \frac{d\varphi\left(x\right)}{\varphi_{0}}\,
    e^{\,d^{D}x\lp x\rp f_{j}
    \left[\varphi\left(x\right);x\right]}}.
\end{split}
\end{equation}

Change spatial hypercube $d^{D}x$ to measure $d\mu(x)$ (for the spatial and momentum space measures we use the same notation). Set source $j(x)=0$, and introduce HS continuous index vectors $\sum\limits_{\sigma}\int d\mu(k)t_{\sigma,k}D_{\sigma,k}(x)\equiv \vec{t}\star\vec{D}(x)$, where star means scalar product of two vectors with uncountable number of components. The expression for GF $\mathcal{Z}$ takes the form (we use operator notation for functions $\exp$ and $\varPsi^{-1}$):
\begin{equation}\label{GFZ_Continuous_Lattice}
    \mathcal{Z}\left[g;d\mu\right]=
    \int\mathcal{D}\sigma\lb t\rb 
    \exp\int d\mu\left(x\right)\ln
    \varPsi^{-1}\int\limits_{-\infty}^{+\infty}
    \frac{d\varphi\left(x\right)}{\varphi_{0}}\,
    \varPsi\left\{e^{-g\left(x\right)
    U\lb\frac{\varphi\left(x\right)}
    {\varphi_{0}}\rb+
    \vec{t}\star\vec{D}\left(x\right)
    \varphi\left(x\right)}\right\}.
\end{equation}

The expression (\ref{GFZ_Continuous_Lattice}) for GF $\mathcal{Z}$ is general. Further calculation is possible only after specifying the theory. In the next subsection, we consider polynomial theory $\varphi_{D}^{4}$

\subsection{Polynomial Theory $\varphi_{D}^{4}$}

Restrict ourselves to the case of the polynomial theory $\varphi_{D}^{4}$. Change variables $\frac{\varphi\left(x\right)}{\varphi_{0}}= \frac{\xi\left(x\right)}{\sqrt[4]{g\left(x\right)}}$ and introduce notation
$\alpha(x)=\phi_{0}\,\vec{t}\star\vec{D}(x)$:
\begin{equation}
\begin{split}
    \mathcal{Z}\left[g;d\mu\right]&=
    \int\mathcal{D}\sigma\lb t\rb 
    \exp\int d\mu\left(x\right)\ln
    \varPsi^{-1}\int\limits_{-\infty}^{+\infty}
    \frac{d\xi\lp x\rp}{\sqrt[4]{g\lp x\rp}}\,
    \varPsi\left\{e^{-\xi^{4}\left(x\right)+
    i\frac{\alpha\left(x\right)}
    {\sqrt[4]{g\lp x\rp}}\xi\lp x\rp}\right\}\\
    &=\int\mathcal{D}\sigma\left[t\right]
    e^{\,\int d\mu\left(x\right)
    \ln\frac{\varepsilon_{1}}{\sqrt[4]
    {g\left(x\right)}}}
    \left\{1+\int\frac{d\mu\left(x\right)}
    {\sqrt{g\left(x\right)}}\,
    \varepsilon_{3}\left(x\right)+
    \mathcal{O}\left[\frac{1}{g}\right]\right\},
\end{split}
\end{equation}
where smooth generator $\varPsi$ is used.

From the third section, first lines of expressions (\ref{For_future_calculations_1}) and (\ref{For_future_calculations_2}), it is known that:
\begin{equation}
    \int\mathcal{D}\sigma\lb t\rb \varepsilon_{3}\left(x\right)= \frac{1}{3}\varepsilon^{2}_{1}- \frac{\varepsilon_{3}}{\varepsilon_{1}}
    \int \mathcal{D}\sigma\lb t\rb
    \alpha^{2}\left(x\right).
\end{equation}

Consider averaging with Gaussian measure $\int \mathcal{D}\sigma\lb t\rb$ in more detail:
\begin{equation}
\begin{split}
    \int\mathcal{D}\sigma\lb t\rb
    \alpha^{2}\left(x\right)&=
    \varphi^{2}_{0}\sum\limits_{\sigma,\,\sigma'}
    \int d\mu\left(k\right)\int d\mu\left(k'\right)
    D_{\sigma,k}\left(x\right)
    D_{\sigma',k'}\left(x\right)
    \underbrace{\int\mathcal{D}\sigma\lb t\rb
    t_{\sigma,k}t_{\sigma',k'}}_{\frac{1}
    {d\mu\left(k'\right)}
    \,\delta_{\sigma,\sigma'}\delta_{k,k'}}\\
    &=\varphi^{2}_{0}\sum\limits_{\sigma}
    \int d\mu\left(k\right)
    D_{\sigma,k}\left(x\right)
    D_{\sigma,k}\left(x\right)=
    \varphi^{2}_{0}L\left(0\right).
\end{split}
\end{equation}

Thus, the ``order parameter'' $\eta=\varphi_{0}^{2}L\lp 0\rp$ appears again, and the final expression for GF $\mathcal{Z}$ coincides with the similar on the discrete lattice of functions, the first line of (\ref{For_future_calculations_3}):
\begin{equation}
    \mathcal{Z}\lb g;d\mu\rb=
    e^{\int d\mu\lp x\rp\ln\frac{\varepsilon_{1}}
    {\sqrt[4]{g\lp x\rp}}}\lc 1+
    \lb\frac{1}{3}\varepsilon_{1}^{2}-
    \frac{\varepsilon_{3}}{\varepsilon_{1}}
    \varphi_{0}^{2}L\left(0\right)\rb
    \int\frac{d\mu\lp x\rp}{\sqrt{g\lp x\rp}}+
    \mathcal{O}\left[\frac{1}{g}\right]\rc.
\end{equation}

In conclusion, let us consider the structure of the higher order terms such as $\int\mathcal{D}\sigma\lb t\rb \alpha^{2}(x)\alpha^{2}(y)$, etc. For the given example:
\begin{equation}
    \int\mathcal{D}\sigma\lb t\rb
    \alpha^{2}\left(x\right)
    \alpha^{2}\left(y\right)=
    \varphi^{4}_{0}\left\{L^{2}\left(0\right)+
    2L^{2}\left(x-y\right)\right\},
\end{equation}
due to (page $192$ in \cite{wipf2012statistical})
\begin{equation}
    \int\mathcal{D}\sigma\lb t\rb
    t_{\sigma_{1},k_{1}}t_{\sigma_{2},k_{2}}
    t_{\sigma_{3},k_{3}}t_{\sigma_{4},k_{4}}=
    \frac{1}{d\mu\left(k_{2}\right)}\,
    \delta_{\sigma_{1},\sigma_{2}}\delta_{k_{1},k_{2}}
    \frac{1}{d\mu\left(k_{4}\right)}\,
    \delta_{\sigma_{3},\sigma_{4}}\delta_{k_{3},k_{4}}+
    \text{permutations}.
\end{equation}

Thus, the results on a discrete lattice of functions (the number of auxiliary variables $t_{s}$ is countable) are reproduced, and the theory considered in the main part of the paper receives one more confirmation.

\subsection{GF $\mathcal{Z}$ in Terms of the Parseval--Plancherel Identity}

In this subsection, we consider the Parseval--Plancherel identity -- the strongest tool that allows to establish different dualities in QFTs at the functional integral level (page $176$ in \cite{Shavgulidze2015}). With the help of the latter, we derive the expression (\ref{GFZ_Continuous_Lattice}) for GF $\mathcal{Z}$ on a continuous lattice of functions. Thus, the expression (\ref{GFZ_Continuous_Lattice}) will receive an independent verification, and therefore the whole theory on a discrete lattice of functions.

First, we rewrite GF $\mathcal{Z}$ (\ref{generation_function_interaction}) as follows:
\begin{equation}\label{Parseval_Plancherel_Identity_1}
    \mathcal{Z}\lb j,g\rb=\int\mathcal{D}\lb\varphi\rb \mathcal{F}\lb\varphi,j\rb
    \mathcal{G}\lb\varphi,j\rb,
\end{equation}
where we split the integrand into two pieces ($\alpha$ and $\beta$ are constants):
\begin{equation}
    \mathcal{F}\lb\varphi,j\rb = e^{-S_{0}\lb\varphi\rb+\alpha\lp j|\varphi\rp},\quad  \mathcal{G}\lb\varphi,j\rb = 
    e^{-S_{1}\lb\varphi\rb+\beta\lp j|\varphi\rp},\quad \alpha+\beta =1.
\end{equation}

Next, we imply the Parseval--Plancherel identity in (\ref{Parseval_Plancherel_Identity_1}). Briefly clarify the derivation: This identity is easily proved, at least, at \emph{the formal calculus level}. One needs to use the formal Fourier transform of the integrands in (\ref{Parseval_Plancherel_Identity_1}), then integrate over the primary field $\varphi$ (such an integral is equal to the delta-functional -- the infinite-dimensional Dirac delta-function), and finally integrate the delta-functional over one of the Fourier transform fields. Thus, we obtain GF $\mathcal{Z}$ as an integral over the remaining Fourier transform field and with Fourier transforms of the functionals introduced in (\ref{Parseval_Plancherel_Identity_1}):
\begin{equation}\label{Parseval_Plancherel_Identity_2}
    \mathcal{Z}\lb j,g\rb= \int\mathcal{D}\lb k\rb \mathcal{\tilde{F}}\lb k,j\rb 
    \mathcal{\tilde{G}}^{*}\lb k,j\rb,
\end{equation}
where
\begin{equation}\label{Parseval_Plancherel_Identity_3}
    \mathcal{\tilde{F}}\lb k,j\rb = 
    \int\mathcal{D}\lb\varphi\rb 
    \mathcal{F}\lb\varphi,j\rb 
    e^{-i\lp k|\varphi \rp}, \quad 
    \mathcal{\tilde{G}}^{*}\lb k,j\rb = \int\mathcal{D}\lb\varphi\rb 
   \mathcal{G}\lb\varphi,j\rb 
   e^{i\lp k|\varphi \rp}.
\end{equation}   

The first one in (\ref{Parseval_Plancherel_Identity_3}) can be easily calculated because it is nothing else but a Gaussian integral: 
\begin{equation}
   \mathcal{\tilde{F}}\lb k,j\rb =
   \int\mathcal{D}
   \lb\varphi\rb e^{-S_{0}\lb\varphi\rb+
   \lp \alpha j-ik|\varphi\rp} = 
   \mathcal{Z}_{0}\,
   e^{\frac{1}{2}\alpha^{2}\lp j|\hat G|j\rp-
   \frac{1}{2}\lp k|\hat G|k\rp-
   i\alpha\lp k|\hat G|j\rp},
\end{equation}
where $\hat G=\hat L^{-1}$ is the Gaussian propagator, and the constant $\mathcal{Z}_{0}=1/\sqrt{\mathrm{Det}\hat L}$ does not depend on fields and sources.

As for $\mathcal{\tilde{G}}^{*}\lb k,j\rb$ it is much more tricky to do but we can calculate it using the developed technique. The explicit expression for $\mathcal{\tilde{G}}^{*}\lb k,j\rb$ reads:
\begin{equation}
    \mathcal{\tilde{G}}^{*}\lb k,j\rb=
    \int\mathcal{D}\lb\varphi\rb 
    e^{-S_{1}\lb\varphi\rb+
    \lp\beta j+ik|\varphi\rp},
\end{equation}
where for further convenience we denote the complete source $J_{k}=\beta j+ik$. Recalling the results of the general theory section, in particular, the expression (\ref{0-norm_general}), the functional $\mathcal{\tilde{G}}^{*}\lb k,j\rb$ can be represented as: 
\begin{equation}
    \mathcal{\tilde{G}}^{*}\lb k,j\rb=
    \exp\int d\mu\lp x\rp\ln\varPsi^{-1}
    \int\limits_{-\infty}^{+\infty} \frac{d\varphi(x)}{\varphi_{0}}\,
    \varPsi\lc e^{\,f_{j,\,k}\lb \varphi\lp x\rp;x \rb}\rc,
\end{equation}
where the function $f_{j,\,k}\lb\varphi;x\rb=-g(x)U\left(\frac{\varphi}
{\varphi_{0}}\right) + J_{k}(x)\varphi$.

Therefore the full expression for GF $\mathcal{Z}$ in terms of the Parseval--Plancherel identity reads:
\begin{equation}\label{Parseval_Plancherel_Identity_4}
\begin{split}
    \mathcal{Z}\lb j,g\rb&=
    \mathcal{Z}_{0}\,
    e^{\frac{1}{2}\alpha^{2}\lp j|\hat G|j\rp}
    \int\mathcal{D}\lb k\rb
    e^{-\frac{1}{2}\lp k|\hat G|k\rp-
    i\alpha\lp k|\hat G|j\rp}\\
    &\times\exp\int d\mu\lp x\rp
    \ln\varPsi^{-1}\int\limits_{-\infty}^{+\infty}
    \frac{d\varphi\lp x\rp}{\varphi_{0}}
    \varPsi\lc e^{-g\lp x\rp 
    U\lb\frac{\varphi\lp x\rp}{\varphi_{0}}\rb+
    \beta j\lp x\rp\varphi\lp x\rp+
    ik\lp x\rp\varphi\lp x\rp}\rc.
\end{split}
\end{equation}

Consider the choice of parameters $\alpha=0,\,\beta=1$. Also make the change of the integration variable (field $k$) as follows: $k=\hat{D}t$, where the operator $\hat{D}$ is defined according to the general theory section as $\hat{G}=\hat{D}^{-1}\hat{D}^{-1}$. The expression (\ref{Parseval_Plancherel_Identity_4}) now reads:
\begin{equation}\label{Parseval_Plancherel_Identity_5}
    \mathcal{Z}\lb j,g\rb= 
    \int\mathcal{D}\sigma\lb t\rb
    \exp\int d\mu\lp x\rp\ln\varPsi^{-1}
    \int\limits_{-\infty}^{+\infty}
    \frac{d\varphi\lp x\rp}{\varphi_{0}}
    \varPsi\lc e^{-g\lp x\rp 
    U\lb \frac{\varphi\lp x\rp}{\varphi_{0}}\rb+
    j\lp x\rp\varphi\lp x\rp +i\hat{D}t\lp x\rp
    \varphi\lp x\rp}\rc,
\end{equation}
where the Gaussian measure (up to a constant) $\int\mathcal{D}\sigma\lb t\rb\equiv
\int\mathcal{D}\lb t\rb\,e^{-\frac{1}{2}\lp t|t\rp}$ is introduced. 

Finally, consider the field $\hat{D}t\lp x\rp$ in (\ref{Parseval_Plancherel_Identity_5}). 
Transform this field as follows:
\begin{equation}\label{Parseval_Plancherel_Identity_6}
    \lb\hat{D}t\rb\lp x\rp=
    \int d^{D}y D\lp y-x\rp t\lp y\rp=
    \int d\mu\lp k\rp D\lp -k\rp t\lp k\rp e^{ikx}=\vec{t}\star\vec{D}\lp x\rp.
\end{equation}

The last equality in the expression (\ref{Parseval_Plancherel_Identity_6}) is valid due to the trigonometric form for the complex- valued exponent. Thus, expressions (\ref{Parseval_Plancherel_Identity_5}) and (\ref{GFZ_Continuous_Lattice}) are coincide (in the latter one can always take into account a non-zero source $j$) and the mathematical technique developed in the present paper passes one more test by an independent verification method.

\section{Conclusions}

In this paper we have studied GF of complete Green functions $\mathcal{Z}$ as a functional of an external source $j$, coupling constant $g$, and spatial measure $d\mu$, both for polynomial and nonpolynomial interaction Lagrangians $U\left(\varphi\right)$. Important properties of the interaction Lagrangian chosen in this paper are: $U\left(-\varphi\right)=U\left(\varphi\right)$ and $U\left(\varphi\rightarrow\infty\right)\rightarrow\infty$ (even and unbounded function). If the Lagrangian was bounded, e.g. $U\left(\varphi\right)=\sin^{4}{\varphi},\, \varphi^{2n}e^{-\varphi^{2}},\, n=2,3,4,\ldots,$ etc. for the analysis of GF $\mathcal{Z}$ or $\mathcal{S}$-matrix, one could use the representation for $\mathcal{Z}$ or $\mathcal{S}$ in terms of the grand canonical partition function \cite{efimov1970nonlocal,efimov1977cmp,efimov1979cmp,efimov1977nonlocal,efimov1985problems,Ogarkov2019}. Such a partition function is classical, i.e. has a finite radius of convergence, series over the interaction constant $g$, and can be explored with statistical physics methods.

We have defined two functional integration measures $\mathcal{D}\left[\varphi\right]$ over the primary field $\varphi$, associated with $1$-norm and $0$-norm, respectively. In the case of $1$-norm, we have introduced an original symbol for measure $\mathcal{D}\left[\varphi\right]$ in terms of a product integral. This way, we have calculated the integral over the primary field $\varphi$ in quadratures, remaining with the integral over the HS only and with complicated integrand. As one can see from the plots, presented in the paper, the integrand is not a positive definite quantity. For this reason, an interesting question of investigating the phase of such an integrand arises.

To overcome this complexity as well as to develop some mathematical intuition about the object obtained, we have proposed the calculation of GF $\mathcal{Z}$ in terms of the inverse coupling constant $1/\sqrt{g}$. Alternatively, various inequalities could be used for analysis: Jensen inequality, H\"{o}lder inequality, etc. Besides, as a necessity we have introduced a spatial measure $d\mu$. In a sense, such a measure is equivalent to a smooth switch on-off of interaction with coupling constant $g$. The physical realization of this measure can be an infinitely small $D$-dimensional hypercube $d^{D}x$. Alternatively, in a real model, spatial measure $d\mu$ can be a part of the problem statement and is determined by physical requirements. This additional degree of freedom is useful in nuclear physics problems.

Further, we have introduced replica-functional Taylor series -- a generalization of a functional Taylor series. This generalization allows to determine $0$-norm in the form in which it is required for the functional integral calculation. A special case of this definition is the definition in terms of the $0$-norm generator $\varPsi$. We have suggested one sharp and two smooth generators $e^{\,i\arg{z}}\min\left\{|z|,1\right\}$, $\arctan(z)$ and $\tanh(z)$, respectively. In the sharp generator case, $0$-norm and $1$-norm coincide. In general, this is also an additional degree of freedom.

We have presented final results for the polynomial theories $\varphi^{2n},\, n=2,3,4,\ldots,$ and for the nonpolynomial theory $\sinh^{4}\varphi$. For the sharp $0$-norm ($1$-norm) in the nonpolynomial case, an interesting dependence on a dimensionless parameter $\eta=\varphi_{0}^{2}L\left(0\right)$ was arisen. This parameter is the ratio of the ultraviolet length parameter from $L\left(0\right)$, and some length parameter from $\varphi_{0}$. The latter in turn determines the scale of the interaction Lagrangian $U\left(\varphi\right)$ and must be considered ultraviolet. Thus, a phase transition in the nonpolynomial theory $\sinh^{4}\varphi$ was found.

An interesting picture shows up: The interaction Lagrangian $U\left(\varphi\right)$ is driven by the ultraviolet length scale, but the spatial measure $d\mu$ is driven by some infrared length scale. Such a hierarchy of length scales forms a nontrivial nonpolynomial field theory \cite{basuev1973conv,basuev1975convYuk}. The critical value of $\eta$ (where $1/\sqrt{g}$ correction changes sign) is $\eta_{c}=1$. Also, we have found a phase transition in the case of a smooth generator $\varPsi$ for both polynomial $\varphi^{4}$ and nonpolynomial $\sinh^{4}\varphi$ theories. We have calculated critical parameters values for two $0$-norms with generators $\arctan(z)$ and $\tanh(z)$ numerically. These results could be useful for effective field theories in nuclear physics.

In conclusion, we give a short list of further possible research directions: nonlocal scalar QFT in curved spacetime, nonlocal scalar QFT in abstract spacetime, nonlocal scalar QFT in spacetime with extra dimensions and compactification, $N$-component model, nonlocal QFT with fermions, nonlocal Yukawa model, GF $\mathcal{Z}$ on different source configurations, different generating functionals, different Green functions families and different composite operators, analytical continuation to Minkowski spacetime, Schwinger--Dyson and Tomonaga--Schwinger equations in nonlocal QFT and iteration procedure for solution, nonlocal QFT methods in stochastic partial differential equations theory, noncommutative nonlocal QFT, etc. Any such a research will necessarily provide answers to the mathematical questions of quantum field theory that are open to date.

\vspace{8pt}
\authorcontributions{M.B.: conceptualization, methodology, formal analysis, investigation, writing--original draft preparation, writing--review and editing; V.A.G.: software, validation, formal analysis, investigation, writing--original draft preparation, visualization; M.G.I.: conceptualization, methodology, formal analysis, investigation, writing--original draft preparation, writing--review and editing; A.E.K.: software, validation, formal analysis, investigation, writing--original draft preparation, visualization; S.L.O.: conceptualization, methodology, formal analysis, investigation, writing--original draft preparation, writing--review and editing, supervision.}

\vspace{8pt}
\funding{This research received no external funding}

\vspace{8pt}
\acknowledgments{The authors are deeply grateful to their families for their love, wisdom and understanding. We are very grateful to Ivan V. Chebotarev for his help with Overleaf, Online LaTeX Editor. 

The authors are grateful to organizers and participants of the International Conference ``Mathematical Physics, Dynamical Systems and Infinite-Dimensional Analysis'', who made valuable comments on the report based on the materials presented in this paper, especially to Grigoriy E. Ivanov, Vsevolod Zh. Sakbaev and Nikolay N. Shamarov. 

S.L.O. is grateful to Mikhail D. Yudaev for helpful discussions of some functional analysis issues. M.B. was supported by The Lynn Bit Foundation. We are grateful to Dr. Les\l{}aw Rachwa\l{}, for the links to the interesting papers on nonlocal QFT \cite{Modesto2014,Modesto2018,Modesto2015}. 

Finally, we express special gratitude to Sergey E. Kuratov and Alexander V. Andriyash for supporting this research at an early stage at the Center for Fundamental and Applied Research (Dukhov Research Institute of Automatics).}

\vspace{8pt}
\conflictsofinterest{The authors declare no conflict of interest.} 

\newpage

\abbreviations{The following abbreviations are used in this manuscript:\\

\noindent 
\begin{tabular}{@{}ll}
FRG &  Functional Renormalization Group\\
GF &  Generating Functional\\
GLT & Ginzburg--Landau Theory\\
HS &  Hilbert Space\\
PT &  Perturbation Theory\\
QCD &  Quantum Chromodynamics\\
QED &  Quantum Electrodynamics\\
QFT &  Quantum Field Theory\\
QG &  Quantum Gravity\\
RG &  Renormalization Group
\end{tabular}}

\reftitle{References}

\end{document}